\documentclass[useAMS,usenatbib]{mn2e}
\usepackage{graphics,rotate}
\def\gtsim{\mathrel{\hbox{\rlap{\hbox{\lower4pt\hbox{$\sim$}}}\hbox{$>$}}}}
\def\lesssim{\mathrel{\hbox{\rlap{\hbox{\lower4pt\hbox{$\sim$}}}\hbox{$<$}}}}

\def\spose#1{\hbox to 0pt{#1\hss}}
\def\approxlt{\mathrel{\spose{\lower 3pt\hbox{$\sim$}}
	\raise 2.0pt\hbox{$<$}}}
\def\approxgt{\mathrel{\spose{\lower 3pt\hbox{$\sim$}}
	\raise 2.0pt\hbox{$>$}}}
\def\approxpropto{\mathrel{\spose{\lower 3pt\hbox{$\sim$}}
	\raise 2.0pt\hbox{$\propto$}}}
\mathchardef\twiddle="2218

\def\multleft#1{\hbox to size{\vbox {\halign {\lft{##}\cr #1}}\hfill}\par}
\def\multright#1{\hbox to size{\vbox {\halign {\rt{##}\cr #1}}\hfill}\par}

\def\today{\ifcase\month\or January\or February\or March\or April\or May\or
      June\or July\or August\or September\or October\or November\or December\fi
      \space\number\day, \number\year}
\def\<{\thinspace}
\def\km{{\rm\thinspace km}}

\def\Mpc{{\rm\thinspace Mpc}}
\def\Msun{\hbox{$\rm\thinspace M_{\odot}$}}

\def\s{{\rm\thinspace s}}
\def\yr{{\rm\thinspace yr}}

\def\kmps{\hbox{$\km\s^{-1}\,$}}

\def\Msunpyr{\hbox{$\Msun\yr^{-1}\,$}}

\def\kmpspMpc{\hbox{$\kmps\Mpc^{-1}$}}

\title{{\it Chandra} observations of the galaxy cluster Abell 1835%
}

\author[Schmidt, Allen \& Fabian]
       {R. W. Schmidt,\thanks{E-mail: rschmidt@ast.cam.ac.uk (RWS),
       swa@ast.cam.ac.uk (SWA), acf@ast.cam.ac.uk (ACF)}
        S. W. Allen and A. C. Fabian\\
        Institute of Astronomy, University of Cambridge, Madingley Road,
	Cambridge CB3 0HA, United Kingdom}
\date{%Accepted .
      Received }

\pagerange{\pageref{firstpage}--\pageref{lastpage}}
\pubyear{}

\begin{document}

\maketitle

\label{firstpage}

\begin{abstract}

\noindent We present the analysis of 30\,ksec of {\it Chandra}
observations of the galaxy cluster Abell 1835. Overall, the X-ray
image shows a relaxed morphology, although we detect substructure in
in the inner 30\,kpc radius. Spectral analysis shows a steep drop in
the X-ray gas temperature from $\sim$12\,keV in the outer regions of
the cluster to $\sim$4\,keV in the core. The {\it Chandra} data
provide tight constraints on the gravitational potential of the
cluster which can be parameterized by a \protect\cite*{Navarro97}
model. The X-ray data allow us to measure the X-ray gas mass fraction
as a function of radius, leading to a determination of the cosmic
matter density of $\Omega_{\rm m}=0.40\pm0.09\,h_{50}^{-0.5}$. The
projected mass within a radius of $\sim$150\,kpc implied by the
presence of gravitationally lensed arcs in the cluster is in good
agreement with the mass models preferred by the {\it Chandra} data.
%The {\it Chandra} data are also consistent with the
%detection of the Sunyaev-Zeldovich effect by
%\protect\citet{Mauskopf00}.
We find a radiative cooling time of the X-ray gas in the centre of
Abell 1835 of about $3{\times}10^{8}$\,yr. Cooling flow model fits to
the {\it Chandra} spectrum and a deprojection analysis of the {\it
Chandra} image both indicate the presence of a young cooling flow
($\sim 6{\times}10^{8}$\,yr) with an integrated mass deposition rate
of $230^{+80}_{-50}\,M_{\odot}$\,yr$^{-1}$ within a radius of
30\,kpc. We discuss the implications of our results in the light of
recent RGS observations of Abell 1835 with {\it XMM-Newton}.

\end{abstract}


\section{Introduction}
\label{intro}

Since its launch in 1999 July, the {\it Chandra} X-ray observatory
\citep{Weisskopf00} has observed a large number of galaxy clusters, 
many of which are amongst the most massive objects of their kind. 
In particular, high resolution, spatially-resolved spectroscopy with the 
Advanced CCD Imaging Spectrometer (ACIS) on {\it Chandra} is leading 
to significant advances in the way we understand these systems.

As illustrated in a recent paper by \citet{Allen01a}, {\it Chandra}
observations enable us to constrain the properties of galaxy clusters
with unprecedented accuracy. {\it Chandra} data permit direct
measurements of the X-ray gas density, temperature and total mass
profiles at a resolution of $\sim 20$\,h$^{-1}_{50}$\,kpc for the most
massive clusters observed at redshifts $z \sim 0.2$, paving the way
for detailed cosmological studies.

In this paper we present {\it Chandra} observations of the galaxy
cluster Abell 1835 at a redshift $z=0.2523$. This cluster is the most
luminous system ($L_{\rm X}=2.7\times\,10^{45}\,$ergs\,s$^{-1}$ in the
0.1-2.4 keV ROSAT band) in the ROSAT Brightest Cluster Sample
\citep{Ebeling98} and has previously been inferred to contain a strong
cooling flow (with a nominal mass deposition rate of $\sim 1000$
\Msunpyr; e.g., \citealt{Allen96}). The central dominant galaxy
exhibits powerful optical emission lines and a strong UV/blue
continuum associated with large amounts of ongoing star formation
\citep{Allen95,Crawford99}. The observed Balmer emission line ratio
also indicates significant intrinsic reddening
\citep[$E(B$-$V)=0.49^{+0.17}_{-0.15}$]{Allen95}. This is consistent
with the detection of 850\,$\mu$m emission from the cD galaxy by
\citet{Edge99}, which those authors attribute to emission from warm
dust heated by the young, hot stars.

The great concentration of mass in the cluster is also revealed by
several features in optical images that can be identified as
gravitationally lensed background objects. In particular, a large arc
was discovered by Edge et al.~(in preparation) to the south-east of
the cD galaxy. \citet{Allen96} used this arc to put constraints on the
cluster mass inside the radius of the arc. We will discuss several
further possibly lensed objects in this study.

The structure of this paper is as follows: Sect.~\ref{observations}
describes the {\it Chandra} observations and Sect.~\ref{imaging} the
basic imaging results.  Sect.~\ref{spectral} discusses the
spatially-resolved spectroscopy of the cluster using the {\it Chandra}
data. Sect.~\ref{massanalysis} presents a detailed mass analysis using
the independent X-ray and lensing data.% and a comparison with
%observations of the Sunyaev-Zeldovich effect in the cluster.
In Sect.~\ref{cflow} we examine the properties of the central cooling
flow and compare our results with those from recent observations made
with the {\it XMM-Newton} X-ray observatory.  We present our
conclusions in Sect.~\ref{conclusions}. All quantities are given
throughout using a cosmology with H$_0=50$\,km\,s$^{-1}$\,Mpc$^{-1}$
and q$_0=\frac{1}{2}$. Unless otherwise noted all error bars are
1\,$\sigma$ (68.3\%) confidence intervals.

\section{Observations}
\label{observations}

{\it Chandra} observed Abell 1835 on 12 December 1999 for a total of
19.6\,ksec. The target was centred in the middle of node 1 of the
back-illuminated ACIS chip S3 (ACIS-S3), near the nominal aim point of
this detector. The detector temperature during the observation was
-110$^{\degr}$\,C. A further 10.7\,ksec {\it Chandra} observation of
Abell 1835 was obtained on 29 April 2000 at a detector temperature of
\mbox{-120$^{\degr}$\,C}. In this second exposure, the cluster was
centred close to the node boundary in the geometrical centre of the
ACIS-S3 chip, in the middle of a group of five dead columns. This has
prevented us from using this second data set for the spectral analysis
presented in this paper.  We have, however, used both data sets to
produce the images in Figs.~\ref{Image2Mpc} and~\ref{Image150kpc}.

\section{Imaging}
\label{imaging}

\begin{figure}
\protect\resizebox{\columnwidth}{!}
%{\includegraphics{Fig/box.eps}}
%{\includegraphics{Fig/1835_2Mpcx2Mpc.eps}}
{\includegraphics{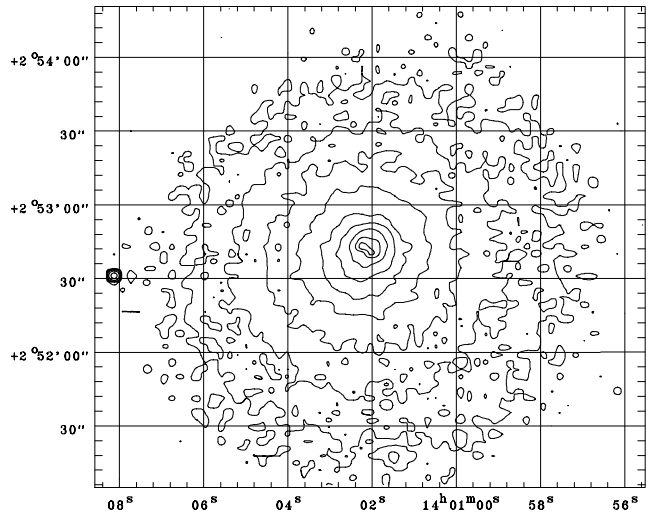}}
\caption{X-ray contours of the central 3.3$\times$3.3\,arcmin
(1x1\,Mpc) of Abell 1835 in the 0.3-7.0\,keV band. The contours were
drawn at 1, 2, 4, 8, 16, 32, 64, 91 and 108 counts per 1x1\,arcsec
pixel. The coordinates are J2000.}
\label{Image2Mpc}
\end{figure}

\begin{figure}
\resizebox{\columnwidth}{!}
%{\includegraphics{Fig/box.eps}}
{\includegraphics{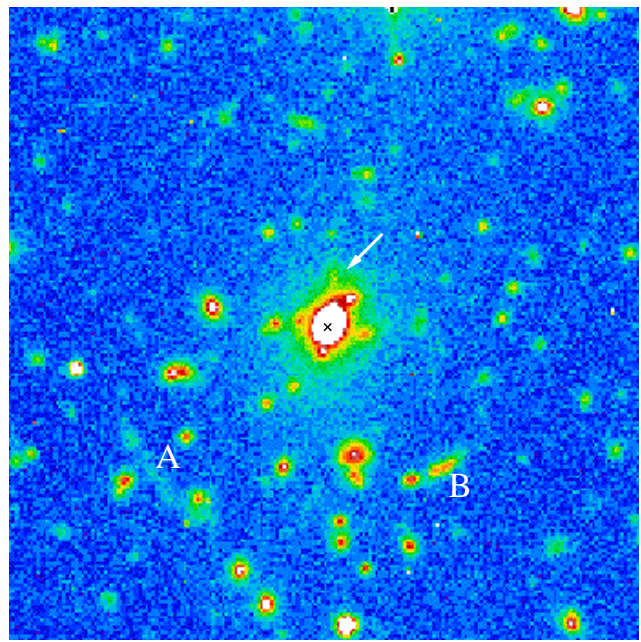}}
\caption{Optical B-band image of the inner 1.5'$\times$1.5'
(0.45$\times$0.45\,Mpc) of Abell 1835. North is up and east is to the
left. The cross marks the galaxy core. The greyscale has been wrapped
once for clarity.}
\label{B-band}
\end{figure}

In Fig.~\ref{Image2Mpc} a contour plot of the 30\,ksec {\it Chandra}
exposure of the inner region of the galaxy cluster in the 0.3-7.0\,keV
energy band is shown with a side length of 3.3 arcminutes (1\,Mpc at
the distance of the galaxy cluster). About 80,000 photons are
collected in this image. For comparison, an archival 20\,min B-band
exposure taken on 27 February 1998 with the Canada France Hawaii
Telescope (CFHT) of the central 1.5'$\times\,$1.5' is shown in
Fig.~\ref{B-band}. In the optical image, the large arc (object A)
discovered by Edge et al.~(in preparation) can be seen to the
south-east of the central cluster galaxy. This arc is thought to be a
gravitationally lensed image of a galaxy far behind Abell 1835. The
presence of this arc, notwithstanding the large X-ray emission,
suggests that Abell 1835 is a very massive galaxy
cluster. \citet{Allen96} estimated the projected mass inside the
radius of the arc to be between 1.4$\times 10^{14}$\,M$_{\odot}$ and
2$\times 10^{14}$\,M$_{\odot}$, if the arc has a redshift greater than
0.7 and is situated on a critical line of the gravitational potential
of the cluster. In Fig.~\ref{B-band} a smaller arc can be seen to the
south-west (object B) of the central cluster galaxy. The arrow to the
north-west of the cD galaxy points to a radial feature which we
discuss in Sect.~\ref{lensing}.

\begin{figure}
\resizebox{\columnwidth}{!}
%{\includegraphics{Fig/box.eps}}
{\includegraphics{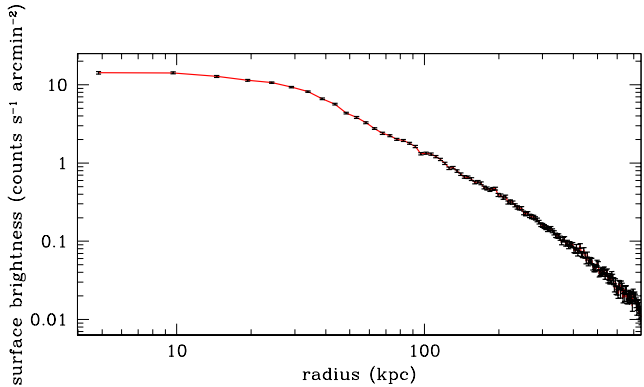}}
\caption{Surface brightness profile of Abell 1835 as a function of
projected separation from the cluster centre. The count rates have
been azimuthally averaged in annuli of width of 1\,arcsec
(4.9\,kpc). The data points are positioned on the abscissa at the
average of the inner and outer radius of the annuli. The data points
are shown with Poisson error bars.}
\label{counts_radius}
\end{figure}

Except for an apparent point source $\sim 1.5$ arcmin to the east
(this is discussed in \citealt{Fabian00a} and \citealt{Crawford00}),
the {\it Chandra} image is dominated by emission from the hot
intracluster gas of the galaxy cluster. Overall, the cluster has a
regular, relaxed morphology. The isophotes outside the core are
elliptical with an axis ratio of minor and major axis $\sim\,0.85$. In
Fig.~\ref{counts_radius}, the average counts per sec per arcsec$^{2}$
as a function of radius are shown as counted in annuli around the
galaxy cluster centre in the 0.3-7.0\,keV band. The profile shows the
steep surface brightness profile typical of cooling flow clusters
\citep[e.g., ][]{Fabian94}. Note, however, that {\it Chandra} detects
a flattening of the profile within $r\sim30\,$kpc.

\label{axisratio}


\begin{figure}
\begin{center}
\resizebox{\columnwidth}{!}
{\includegraphics{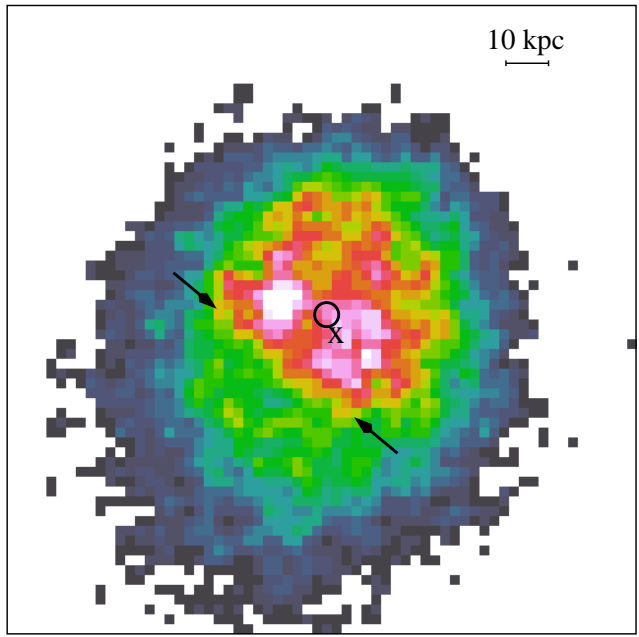}}
\caption{X-ray image of the central 30''$\times$30''
(150\,kpc$\times$150\,kpc) of Abell 1835 in the 0.3-7.0\,keV
band. North is up and east is to the left. The image was smoothed with
a Gaussian with a full-width at half-maximum of 0.8\,arcsec. The
pixels are 0.5x0.5\,arcsec. Only pixels with at least 6 counts are
plotted. The circle marks the geometrical centre of the cluster
mentioned in the text. The cross marks the absolute position of the
galaxy core (see Fig.~\ref{B-band}) determined by \citet{Allen96} when
plotted in the coordinates given by {\it Chandra}. The uncertainties
associated with the galaxy position and the {\it Chandra} astrometry
are $\sim\,1$\,arcsec. The two arrows point at the small emission
front extending between them. The greyscale has been wrapped once for
clarity.}
\label{Image150kpc}
\end{center}
\end{figure}

In Fig.~\ref{Image150kpc} the central 30''$\times$30'' (150\,kpc side
length) of Fig.~\ref{Image2Mpc} are shown. The image was smoothed with
a Gaussian kernel with a full-width at half-maximum of 1.6 pixels (0.8
arcsec).
% 1 sigma=0.7 pixels
In the inner 20\,arcsec (100\,kpc) of the cluster {\it Chandra}
resolves substructure in the X-ray emission. There are a couple of
main features visible: a well-defined peak just east of the
geometrical centre (denoted by the black ring; J2000.0 coordinates
given by {\it Chandra}: RA 14:01:02.07, DEC +2:52:43.2) of the
cluster, a more inhomogeneous structure to the south-west of the
geometrical centre, between these two regions an elongated strip with
a reduced count rate, and a small emission front to the south-east of
the cluster core (between the arrows in Fig.~\ref{Image150kpc}). The
brightness gradient towards the north-west is steeper than towards the
south-east. These features may be the sign of a merging subclump. The
number of counts per half-arcsecond pixel in the brightest regions in
the combined 30\,ksec of our data is between 30 and 40. This means
that the detailed structures need to be analysed with care because the
Poisson error is still of the order of 20\%. The main features,
however, are robustly detected and indicate the presence of variations
of the X-ray emission or absorption in the core of Abell 1835 on
scales of several kiloparsecs.  Comparison with the optical image
(Fig.~\ref{B-band}) shows that the emission inhomogeneities are
situated inside the central cluster galaxy.

\section{Spatially resolved spectral analysis}
\label{spectral}

\subsection{Data extraction}
\label{dataextraction}

We extract spectra from selected regions of the longer 19.6 ksec
exposure {\it Chandra} data set using the CIAO software package
distributed by the {\it Chandra} X-ray Observatory Centre (CXC, web
site: cxc.harvard.edu).

We only work with photons detected in the ACIS-S3 chip,
which is also known as ACIS chip 7. This is a back-illuminated chip
and does not suffer from the degradation of energy resolution that
affects the front-illuminated ACIS chips on the {\it Chandra}
observatory. All significant point sources were masked out and
excluded from the analysis.

Response matrix files (RMF) and ancillary response files (ARF) were
produced using a PERL script generously provided to us by Roderick
Johnstone. This script combines individual RMFs and ARFs for each
32x32 pixel region on the ACIS-S3 chip (generated using the mkrmf tool
in the CIAO software suite available from the CXC website) in a photon
number-weighted manner.

Using the task grppha from the FTOOLS software package provided by the
NASA High Energy Astrophysics Science Archive Research Center
(HEASARC, FTOOLS web site:
heasarc.gsfc.nasa.gov/\linebreak[0]lheasoft/\linebreak[0]ftools/) we
have grouped the extracted spectra so that we have at least 20 counts
in a bin corresponding to a certain range of energy channels. We can
then use Poisson error bars to account for the uncertainty in the
number of photons per bin.

We have have generated background spectra using Maxim Markevitch's
method and program that is available from the CXC web site. We extract
the background data from the same regions on the chip as the spectra
extracted from the Abell 1835 data set. We also scanned the light
curve of the Abell 1835 data for flares or background activity using
the tool lc\_clean provided by Markevitch.  The light curve is stable
throughout the 19.6 ksec exposure.

\subsection{Colour profile analysis}

\begin{figure}
\begin{center}
\rotatebox{270}{\resizebox{!}{\columnwidth}
{\includegraphics{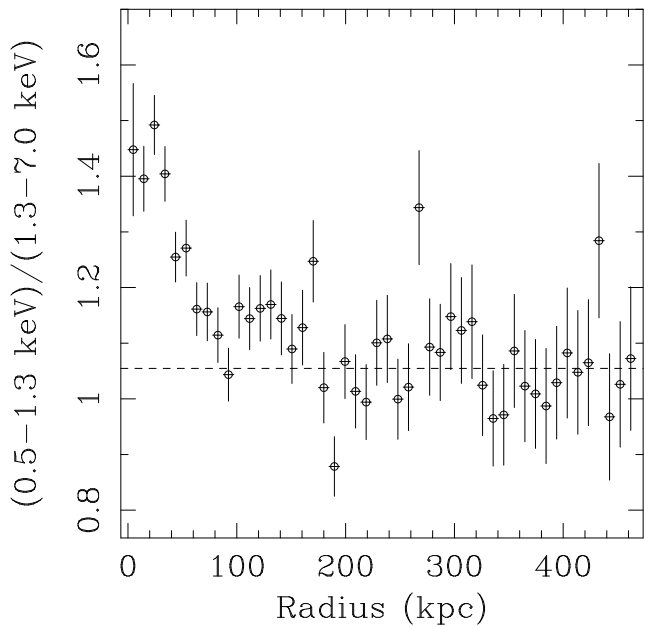}}}
\caption{X-ray colour profile of the central 1.7\,arcmin (500\,kpc) of
Abell 1835. In this figure the ratio of the azimuthally averaged
counts in the $0.5-1.3$ and $1.3-7.0$ keV bands ($0.6-1.6$ and
$1.6-8.8$ keV in the rest frame of the source) is plotted as a
function of the radius of the annuli.
}
\label{colourprofile}
\end{center}
\end{figure}

In order to illustrate the spectral energy distributions from
different regions in Abell 1835 we have firstly constructed an X-ray
colour profile. Two separate images were created in the energy bands
$0.5-1.3$ and $1.3-7.0$ keV ($0.6-1.6$ and $1.6-8.8$ keV in the rest
frame of the source) with a 1.97 arcsec (4 raw detector pixels) pixel
scale. These soft and hard X-ray images were background subtracted and
flat fielded (using the exposure map tools available on the CXC web
site and taking full account of the spectral energy distributions of
the detected photons). Azimuthally-averaged surface-brightness
profiles for the cluster were then constructed in each energy band,
centred on the geometrical centre of the X-ray emission. The colour
profile formed from the ratio of the surface brightness profiles in
the soft and hard bands is shown in Fig.~\ref{colourprofile}.

From examination of Fig.~\ref{colourprofile} we see that at large
radii the observed X-ray colour ratio is approximately constant, with
a mean value of $1.05\pm0.03$ ($1\sigma$ error determined from a fit
to the data between radii of $200-400$ kpc). Within a radius of
$\sim$30\,arcsec (150\,kpc), however, the colour ratio rises,
indicating the presence of cooler gas.

\subsection{Single-temperature model}
\label{singleT}

We investigate now the temperature profile that causes the colour
profile shown in Fig.~\ref{colourprofile}. To do this we have
extracted spectra from annuli around the geometrical centre (marked by
the black circle in Fig.~\ref{Image150kpc}) of the X-ray emission with
photon energies in the 0.5-7.0\,keV band. We use a slightly smaller
energy band for the spectral analysis than we used for the images in
order to get robust answers at soft energies. The annuli have widths
of 5, 10, 25, 50 and 100\,arcsec (25, 50, 125, 250 and 500\,kpc,
respectively) and provide sufficient counts (see
Tab.~\ref{mekal_tab_110}) to study the spectral properties of the
X-ray gas.

We use the software package XSPEC \citep{Arnaud96} to model the
spectra with the MEKAL plasma emission model
\citep{Kaastra93,Liedahl95} multiplied by the PHABS
\citep{Balucinska-Church92} photoelectric absorption model
\begin{equation}
{\rm MODEL}_1 = {\rm PHABS}\,(N_{\rm H})\,\times\,{\rm
MEKAL}\,(T;Z;K).
\label{isothermal}
\end{equation}
The free model parameters are indicated in brackets: the absorbing
equivalent hydrogen column density $N_{\rm H}$, the gas temperature
$T$, the metallicity $Z$ (in units of solar abundance), and a
normalization factor K.

An account of the relevant parameters and the model fits is given in
Table~\ref{mekal_tab_110}. All our fits are statistically acceptable,
with reduced $\chi^{2}$ values of around or slightly above
unity. Since the absorbing column density and the metallicity are not
well constrained in the outer annuli, we have also carried out a
second analysis in which we grouped the data into three larger
regions. This procedure eliminates free parameters and thus yields
more robust estimates. The results are shown in
Table~\ref{mekal_binned}.

It can be seen from Table~\ref{mekal_binned} that the spectra from
annuli with inner radii larger than 30\,arcsec (150\,kpc) are
consistent with Galactic absorption \citep{Dickey90}. We have thus
repeated the isothermal model fits according to eq.~(\ref{isothermal})
for these annuli with fixed Galactic absorption. The corresponding
best-fit models are also given in Table~\ref{mekal_tab_110}. For the
outermost annuli, this leads to a smaller predicted temperature with
smaller uncertainties.

\begin{figure}
\begin{center}
\resizebox{\columnwidth}{!}
{\includegraphics{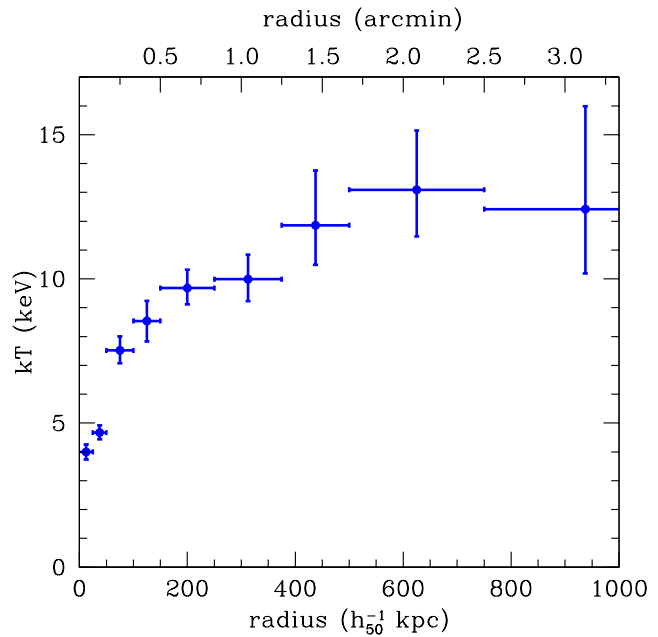}}
\caption{Temperature kT as a function of radius $r$ from the cluster
centre out to 3.3\,arcmin (1\,Mpc). For data points at radii
$\geq30$\,arcsec (150\,kpc), we have plotted the temperatures derived
using models with the absorption fixed at the Galactic value.}
\label{kT}
\end{center}
\end{figure}

In Fig.~\ref{kT} the results from the model fitting for the ambient
gas temperature $T$ are plotted. For annuli with inner radii
$\geq$30\,arcsec (150\,kpc), we have fixed the absorbing column
density to the Galactic value. The best-fit temperatures reveal a
strong drop from $kT\sim\,12$\,keV in the outer regions of the cluster
down to 4\,keV in the centre of the cluster. We note that the photons
in the annuli have been emitted in a hollow cylinder with a
cross-section given by the annulus. The measured temperatures are fits
to the spectrum from this whole region and thus have to be regarded as
average (emission weighted) temperatures.

\begin{table*}
\caption{Isothermal model fits. The model described in
eq.~(\ref{isothermal}) with the free parameters equivalent hydrogen
column density $N_{\rm H}$, temperature T and metallicity $Z$ was
fitted to spectra extracted from annuli around the geometrical centre
of Abell 1835 with inner radii $r_1$ and outer radii $r_2$. For annuli
with $r_1\geq250\,$kpc (50\,arcsec) we have modelled the X-ray spectra
from the annuli both with free equivalent hydrogen column density
$N_{\rm H}$ (top block of models), and with fixed Galactic absorption
\citep[bottom block of models]{Dickey90}. Since neither the
metallicity nor the absorbing column density can be smaller than zero,
the lower limit has always been fixed accordingly for these
parameters. In columns 6 and 7 we list the total number of counts and
the expected number of background counts in the 0.5-7.0\,keV band. In
the last three columns the $\chi^2$ values, the number of degrees of
freedom (DOF) and the reduced $\bar{\chi}^2=\chi^2$/DOF are given.
%, where $\chi^2= \sum_{\,\rm spectrum}\,\left[ (y_{\,\rm obs} - y_{\,\rm
%model})/\sigma_{\,\rm obs} \right]^2.$ $y_{\,\rm model}$ are the
%modelled spectral bins and $y_{\,\rm obs}$ are the observed spectral
%bins with their error bars $\sigma_{\,\rm obs}$. The number of degrees
%of freedom is defined as the difference between the number of spectral
%bins and the number of free model parameters.
}
\label{mekal_tab_110}
\begin{center}
\begin{minipage}{150mm}
\begin{center}
\begin{tabular}{@{}rrrrrrllll@{}}
$r_1$ (kpc) & $r_2$ (kpc) & kT (keV) & $Z$ (solar) & $N_{\rm H}$
(10$^{20}$\,cm$^{-2}$) & counts & exp. background & $\chi^2$ & DOF & $\bar{\chi}^2$ \\
\hline
0 & 25 & 4.0$^{+0.3}_{-0.3}$ & 0.24$^{+0.07}_{-0.07}$       & 3.93$^{+0.82}_{-0.82}$ & 4791 & 2.4 & 148.1 & 122 & 1.21\\
25 & 50 & 4.7$^{+0.2}_{-0.2}$ & 0.31$^{+0.05}_{-0.06}$      & 3.97$^{+0.66}_{-0.64}$ & 8300 & 8.5 & 218.1 & 161 & 1.35\\
50 & 100 & 7.5$^{+0.5}_{-0.4}$ & 0.28$^{+0.06}_{-0.07}$     & 3.27$^{+0.58}_{-0.55}$ & 10776 & 27.0 & 233.8 & 193 & 1.21\\
100 & 150 & 8.5$^{+0.7}_{-0.7}$ & 0.39$^{+0.10}_{-0.10}$    & 2.29$^{+0.66}_{-0.65}$ & 7680 & 51.2 & 186.3 & 172 & 1.08\\
150 & 250 & 9.1$^{+0.8}_{-0.7}$ & 0.35$^{+0.08}_{-0.09}$    & 3.00$^{+0.60}_{-0.58}$ & 10527 & 138.7 & 203.6 & 195 & 1.04\\
250 & 375 & 8.0$^{+0.8}_{-0.6}$ & 0.30$^{+0.09}_{-0.10}$    & 4.75$^{+0.75}_{-0.74}$ & 7838 & 262.6 & 177.0 & 164 & 1.08\\
375 & 500 & 12.5$^{+3.7}_{-1.9}$ & 0.49$^{+0.27}_{-0.24}$   & 1.86$^{+0.96}_{-1.14}$ & 5466 & 378.0 & 135.6 & 124 & 1.09\\
500 & 750 & 16.9$^{+5.1}_{-3.2}$ & 0.20$^{+0.45}_{-0.20}$   & 0.07$^{+2.11}_{-0.07}$ & 7003 & 1123.4 & 159.2 & 169 & 0.94\\
750 & 1250 & 26.0$^{+2.8}_{-11.2}$ & 0.00$^{+0.52}_{-0.00}$ & 0.00$^{+0.29}_{-0.00}$ & 8322 & 3075.7 & 198.0 & 199 & 1.00\\
\\
150 & 250 & 9.7$^{+0.6}_{-0.6}$ & 0.35$^{+0.09}_{-0.09}$   & 2.30 & 10527 & 138.7 & 205.0 & 196 & 1.05\\
250 & 375 & 10.0$^{+0.8}_{-0.8}$ & 0.32$^{+0.12}_{-0.12}$  & 2.30 & 7838 & 262.6 & 188.1 & 165 & 1.14\\
375 & 500 & 11.9$^{+1.9}_{-1.4}$ & 0.48$^{+0.26}_{-0.23}$  & 2.30 & 5466 & 378.0 & 135.8 & 125 & 1.09\\
500 & 750 & 13.1$^{+2.1}_{-1.6}$ & 0.22$^{+0.21}_{-0.12}$  & 2.30 & 7003 & 1123.4 & 162.3 & 170 & 0.95\\
750 & 1250 & 12.5$^{+3.5}_{-2.4}$ & 0.23$^{+0.36}_{-0.23}$ & 2.30 & 8322 & 3075.7 & 207.6 & 200 & 1.04\\
\end{tabular}
\end{center}
\end{minipage}
\end{center}
\end{table*}

\begin{table*}
\caption{Model parameters for isothermal models using combined sets of
data. The spectra from annuli around the cluster centre as used in
Table~\ref{mekal_tab_110} have been fitted simultaneously using the
same absorbing column density and metallicity for all annuli in the
range $r_1\rightarrow\,r_2$ and by allowing for different temperatures
in each of the annuli of Table~\ref{mekal_tab_110}.  The last three
columns, as in Table~\ref{mekal_tab_110}, describe the number of
degrees of freedom and the quality of the fits.}
\label{mekal_binned}
\begin{tabular}{@{}crrrrrrclll@{}}
&$r_1$ (kpc) & $r_2$ (kpc)
& $Z$ (solar) & $N_{\rm H}$& $\chi^2$ & DOF & $\bar{\chi}^2$\\
& & & & & & (10$^{20}$\,cm$^{-2}$) &  (10$^{20}$\,cm$^{-2}$)\\
\hline
& 0 & 50 & 0.29$^{+0.05}_{-0.04}$ & 3.94$^{+0.51}_{-0.51}$ & 367.0 &
285 & 1.29\\
%-110$^{\circ}$
&50 & 150
& 0.31$^{+0.06}_{-0.05}$ & 2.88$^{+0.44}_{-0.43}$
& 422.0 & 367 & 1.15\\
& 150 & 1250 & 
0.31$^{+0.07}_{-0.06}$ & 2.55$^{+0.38}_{-0.36}$
& 898.9 & 857 & 1.05
\end{tabular}
\end{table*}

\begin{figure}
\begin{center}
\resizebox{\columnwidth}{!}
{\includegraphics{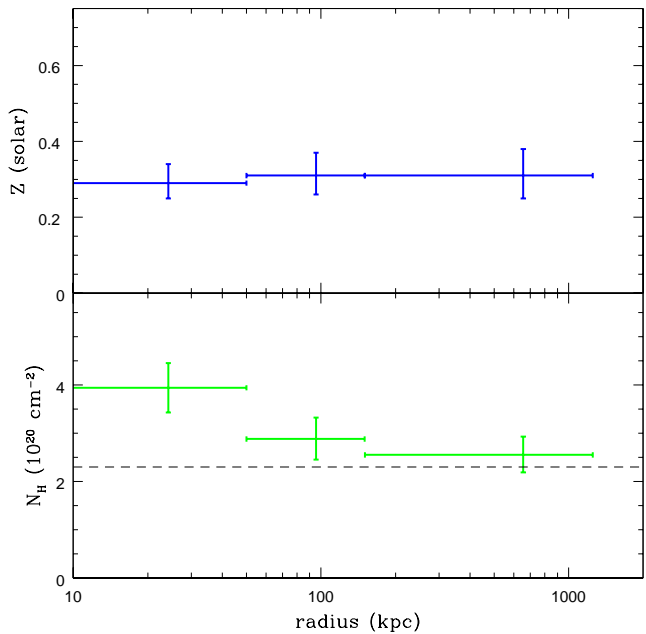}}
\caption{As in Fig.~\ref{kT}, but for metallicity $Z$ (in units of
solar abundance) and photoelectric absorption column density $N_{\rm
H}$ as determined from the combined fits in Table~\ref{mekal_binned}
to X-ray spectra extracted from annuli at several radii. In the lower
panel, the Galactic absorption value from H\,{\sevensize\bf I} studies
is indicated by the dashed line.}
\label{a_nH}
\end{center}
\end{figure}

We have plotted the metallicity and absorption results of
Table~\ref{mekal_binned} in Fig.~\ref{a_nH}. The metallicity appears
approximately constant with radius. We find tentative evidence that
the isothermal model requires excess absorption above the Galactic
value in the inner 100\,kpc.

As an aside, we mention that if we apply model~(\ref{isothermal}) to
the spectrum extracted from a circle with a radius of 30\,arcsec
(150\,kpc) and leave the Ni abundance as an additional free parameter,
we find a Ni abundance of
%$7.8\pm3.0$\,times
$2.3\pm0.8$\,times solar in this region. This relatively high Ni
abundance is relevant for models of Type Ia supernovae and their
enrichment of the intergalactic medium \citep[e.g., ][]{Dupke00}.

\subsection{Spectral deprojection analysis}

The results discussed in the previous section are based on the
analysis of projected spectra. In order to determine the effects of
projection, we have also carried out a deprojection analysis of the
{\it Chandra} spectra using the method described by
\citet{Allen01a}. This method decomposes the observed annular spectra
(Table~\ref{mekal_tab_110}) into the contributions from the X-ray gas
emission from nine spherical shells.

%Spherical symmetry is assumed.

\begin{figure}
\begin{center}
\rotatebox{270}{\resizebox{!}{\columnwidth}
{\includegraphics{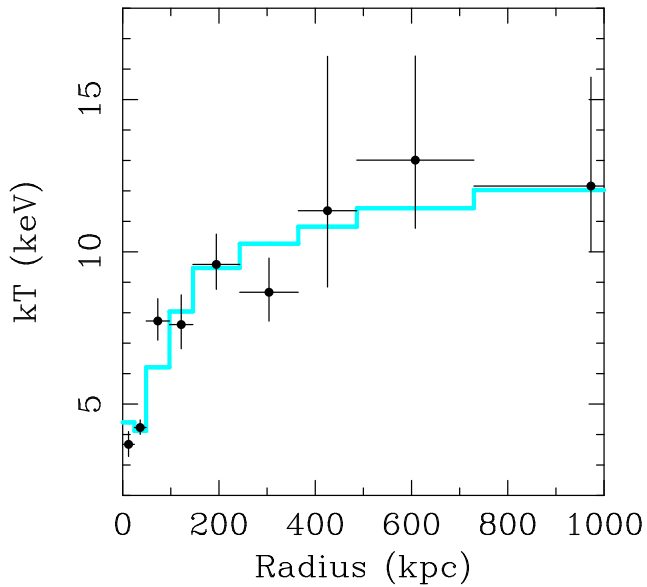}}}
\caption{Deprojected temperature profile of Abell 1835. The solid line
is the best-fit temperature profile determined in
Sect.~\protect\ref{massmodel}.}
\label{kt_deprojected}
\end{center}
\end{figure}

The data for all nine annular spectra were fitted simultaneously in
order to correctly determine the parameter values and confidence
limits. The spectral model used therefore has $2\,n+1$ free parameters
(where $n=9$ is the number of annuli), corresponding to the
temperature and emission measure in each spherical shell and the
overall emission-weighted metallicity (the metallicity is linked to
take the same value at all radii, yielding a metallicity
$Z=0.32^{+0.03}_{-0.04}$). The absorbing column density was fixed at
the Galactic value. The temperature profile determined with the
spectral deprojection code is shown in Fig.~\ref{kt_deprojected}.

\section{X-ray mass analysis} 
\label{massanalysis}

\subsection{The mass model}
\label{massmodel}

The observed X-ray surface brightness profile
(Fig.~\ref{counts_radius}) and deprojected X-ray gas temperature
profile (Fig.~\ref{kt_deprojected}) may together be used to determine
the X-ray gas mass and total mass profiles in the cluster. For this
analysis, we have used an enhanced version of the deprojection code
described by \citet*{White97} and have followed a similar method to
that described by \citet*{Allen01a}.
%We have performed the analysis using only the 20\,ksec of data from the
%$-110^{\circ}$\,C data set.
Spherical symmetry and hydrostatic equilibrium are assumed.

A variety of simple parameterizations for the cluster mass
distribution were examined, to establish which could provide an
adequate description of the {\it Chandra} data. We find that the {\it
Chandra} data can be modelled using a \citet{Navarro97} (NFW) mass
model
\begin{equation}
\rho(r) = {{\rho_{\rm crit}\,\delta_{\rm c}} \over {  ({r/r_{\rm s}}) 
\left(1+{r/r_{\rm s}} \right)^2}},
\end{equation}
where $\rho(r)$ is the mass density, $\rho_{\rm crit} = 3H(z)^2/ 8 \pi
G$ is the critical density for closure of the universe ($H(z)$ is the
Hubble constant and G is the gravitational constant) and
\begin{equation}
\delta_{\rm c} = \frac{200}{3} \frac{ c^3}{ {{\rm
ln}(1+c)-{c/(1+c)}}},
\end{equation}
where $c$ is the concentration parameter. The normalization of the
mass profile may also be expressed in terms of an effective velocity
dispersion, $\sigma = \sqrt{50} H(z) r_{\rm s} c$.

We step the mass model parameters $r_{\rm s}$ and $\sigma$ through a
regular grid of values. (The scale radius $r_{\rm s}$ was stepped
between 0.1\,Mpc and 2.5\,Mpc and the effective velocity dispersion
$\sigma$ between 500\,km\,s$^{-1}$ and 2500\,km\,s$^{-1}$ on a grid of
81$\times$81 models.) Given the observed surface brightness profile
and a particular parameterized mass model, the deprojection code is
used to predict a temperature profile of the X-ray gas (we use the
median model temperature profile determined from 100 Monte-Carlo
simulations) which is then compared for each mass model with the
observed, deprojected temperature profile (Fig.~\ref{kt_deprojected}).

We have determined the goodness of fit of each simulated temperature
profile by calculating the sum
\begin{equation}
\chi^2 = \sum_{\,\rm all\,radial\,bins}\,\left( \frac{T_{\,\rm obs} -
T_{\,\rm model}}{\sigma_{\,\rm obs}} \right)^2,
\end{equation}
where $T_{\,\rm obs}$ is the observed deprojected temperature profile
shown in Fig.~\ref{kt_deprojected} and $T_{\,\rm model}$ is the model
temperature profile, rebinned to the same spatial scale using an
appropriate flux weighting.

Fig.~\ref{kt_deprojected} shows the X-ray gas temperature profile
implied by the best-fitting NFW mass model ($\chi^2=11.9$ for $r_{\rm
s}=0.64^{+0.21}_{-0.12}$\,Mpc,
$\sigma=1275^{+150}_{-100}\,$km\,s$^{-1}$ and $c=4.0^{+0.54}_{-0.64}$)
overlaid on the deprojected temperature profile. The $\chi^2$ value is
mostly due to the differences between model and data close to the
cluster centre, and may at least in part be attributed to the
asymmetric substructure seen in Fig.~\ref{Image150kpc}.

\begin{figure}
\begin{center}
{\resizebox{\columnwidth}{!}
{\includegraphics{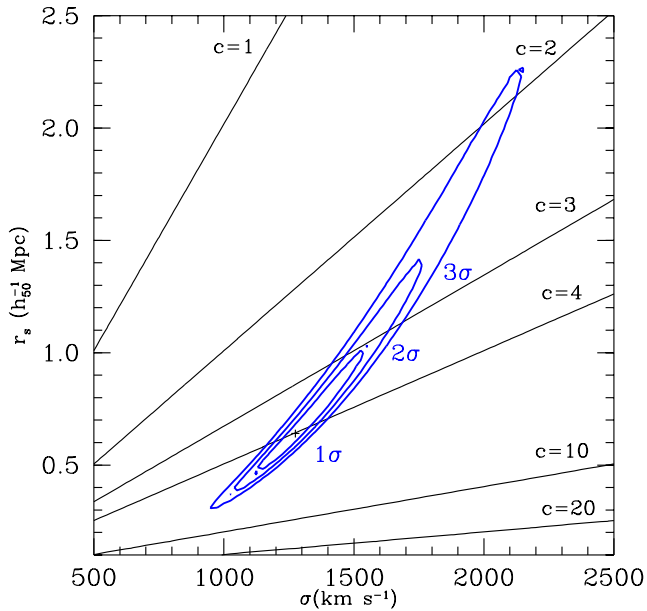}}}
\caption{Contour plot of the 68.3\% ($1\sigma$), 95.4\% ($2\sigma$)
and 99.73\% ($3\sigma$) formal confidence contours on the NFW scale
radius $r_{\rm s}$ and effective velocity dispersion $\sigma$. For
this, a grid of $81\times81$ mass models was graded using the $\chi^2$
statistic by comparing the predicted temperature profile with the
deprojected {\it Chandra} temperature profile. Also shown for
reference are the isocontours of the NFW concentration parameter
$c$. The minimum $\chi^2$ model with $\chi^2=11.9$ is marked by a plus
sign.}
\label{grid}
\end{center}
\end{figure}

\begin{figure}
\begin{center}
{\resizebox{\columnwidth}{!}
{\includegraphics{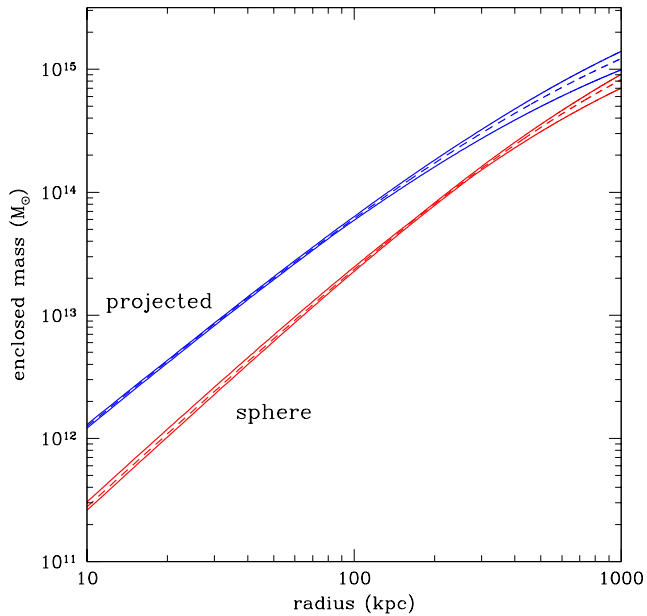}}}
\caption{$1\,\sigma$ range and average value (dashed line) of the
enclosed mass inside a sphere of radius $r$ and the total projected
mass inside a radius $r$.}
\label{massprof}
\end{center}
\end{figure}

Fig.~\ref{grid} shows a contour plot of the $\chi^2$ values calculated
for the different NFW models in the parameter space. The minimum
$\chi^2$ value is marked by a plus sign. Three contours have been
drawn at differences of $\Delta\chi^2=$ 2.3, 6.17 and 11.8,
corresponding to formal confidence intervals of 68.3\% (1\,$\sigma$),
95.4\% (2\,$\sigma$) and 99.73\% (3\,$\sigma$) for the two parameters
\citep[e.g., ][page 661]{Press92}. Fig.~\ref{massprof} shows the
$1\,\sigma$ range of enclosed masses for models around the best-fit
model both for the total mass inside a sphere of radius $r$ and for
the total projected mass inside a radius $r$ (using the mass formula
by \citealt{Bartelmann96}). The mass inside a sphere with the virial
radius $r_{200}=c\times{r_{\rm s}}$ for the best-fit model is
$M_{200}=2.7^{+0.5}_{-0.7}\times10^{15}\,M_{\odot}\,$yr$^{-1}$.

We note that it is also possible to carry out
this analysis with a nonsingular isothermal mass density profile
\begin{equation}
\rho(r)=\frac{\sigma_{\rm iso}^2}{2\,\pi\,{\rm
G}}\,\frac{1}{r^2+r_{\rm c}^2},
\label{isoc}
\end{equation}
where $\sigma_{\rm iso}$ is the velocity dispersion and $r_{\rm c}$ is
the core radius. This yields parameters $r_{\rm
c}=65.0^{+5.0}_{-10.0}\,$kpc and $\sigma_{\rm
iso}=1280\pm60\,$km\,s$^{-1}$. The fit is slightly better,
$\chi^2=6.8$, but the conclusions we can draw from both profiles
regarding the total mass and X-ray gas profiles are very similar.

\subsection{Lensing predictions from the X-ray mass models}
\label{lensing}

\begin{figure*}
\begin{minipage}{\columnwidth}
\begin{center}
\resizebox{\columnwidth}{!}
{\includegraphics{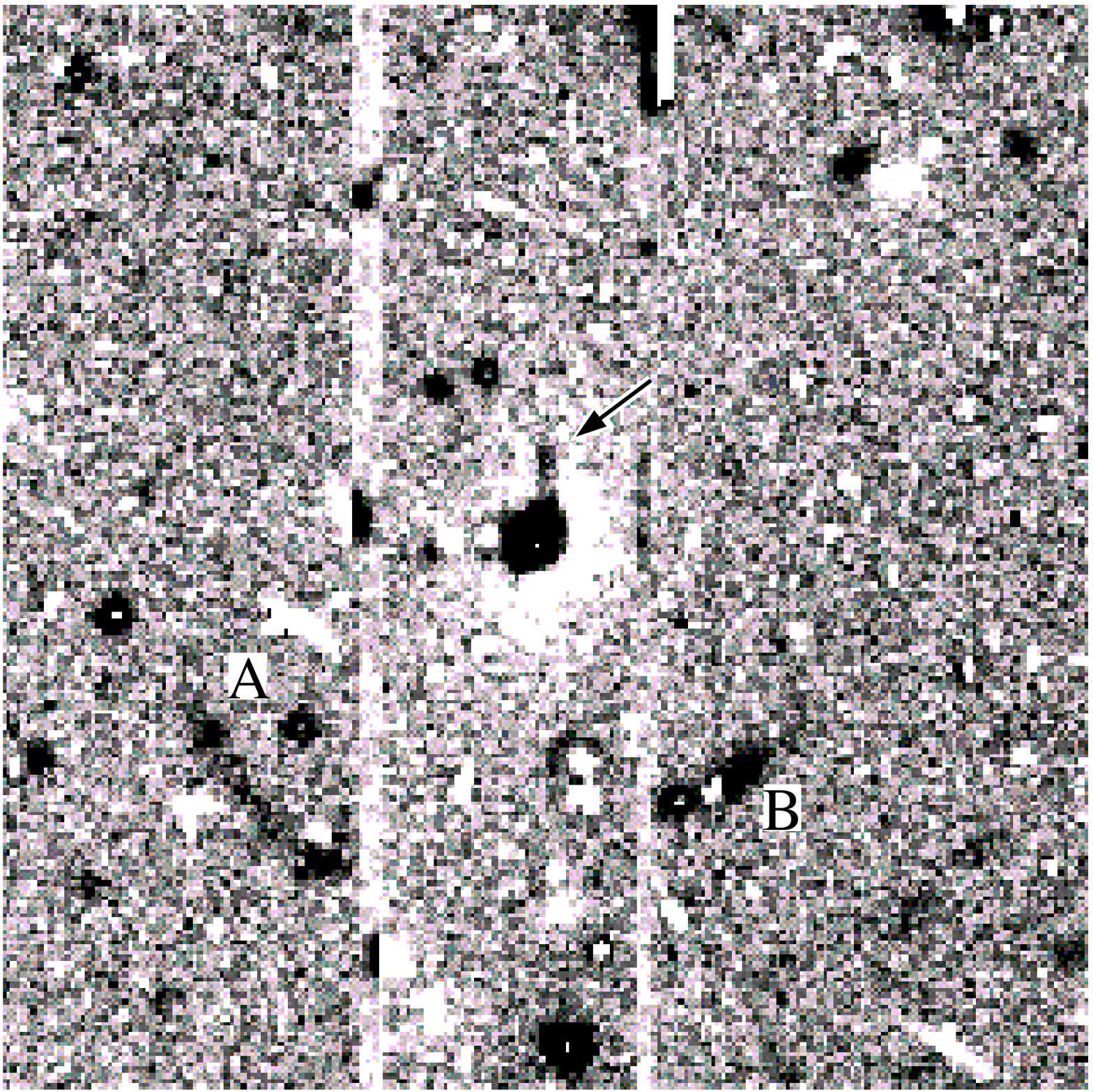}}
\end{center}
\end{minipage}
\hfill
\begin{minipage}{\columnwidth}
\begin{center}
\resizebox{\columnwidth}{!}
{\includegraphics{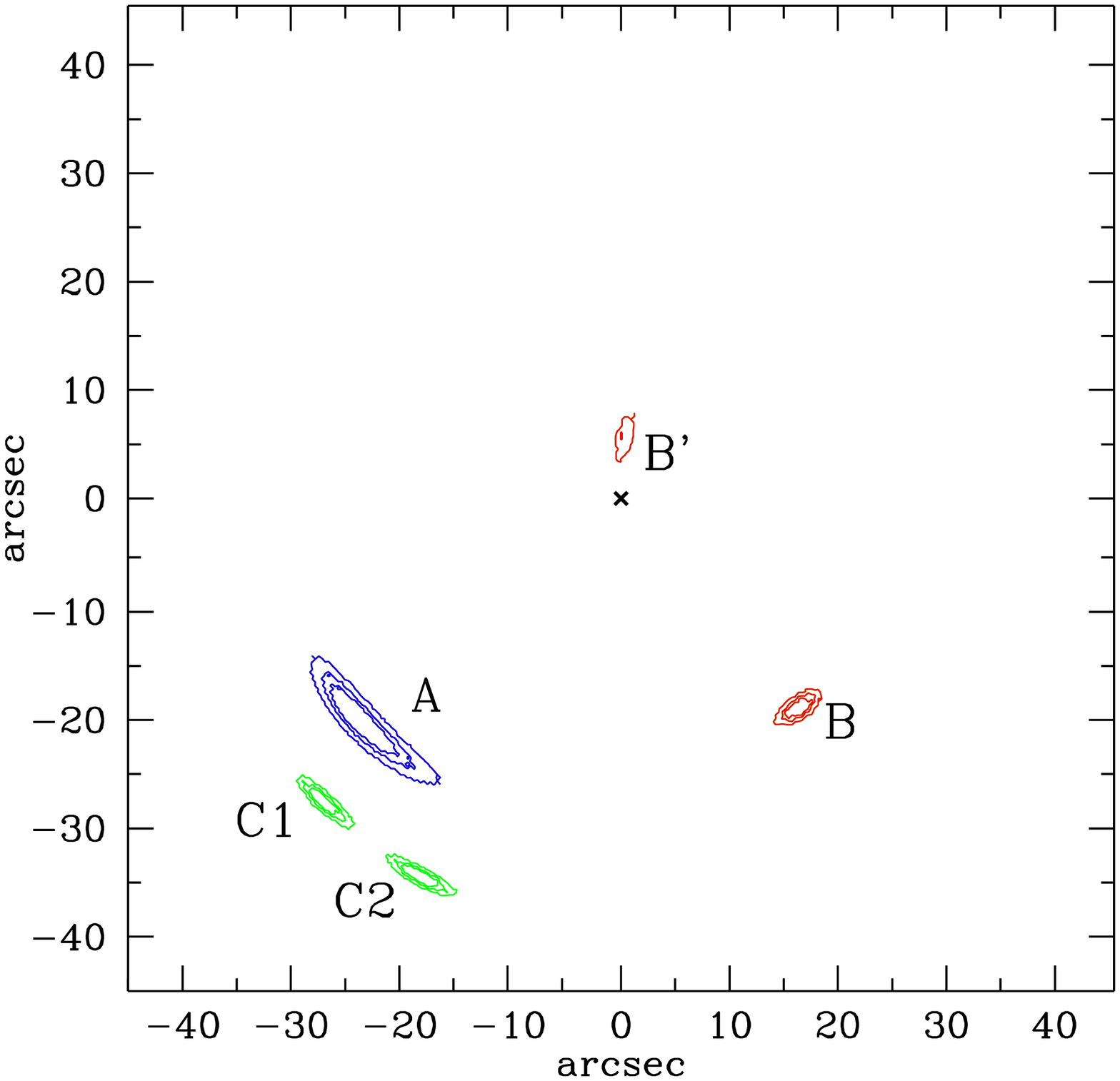}}
\end{center}
\end{minipage}
\caption{{\bf (left panel)} B-R colour difference image of the central
1.5$\times$1.5\,arcmin of Abell 1835 from archival CFHT data. North is
up and East is to the left. The greyscale-contrast has been stretched
to show blue objects in black and red objects in white. The two
vertical features to the left and right of the central cD galaxy are
due to bleeding CCD columns in the R-band image that were caused by
bright stars to the north of the field. We have replaced the pixel
values in these columns by the median background value. Features of
particular interest in this image are the very blue central part of
the cD galaxy and the two arcs to the south east (A) and south west
(B) of the galaxy. We also note that to the north of the cD galaxy a
straight blue feature is seen (arrow). {\bf (right panel)} Contour
plot of gravitationally lensed images by an elliptical NFW profile
with a concentration parameter $c=3.3$, scale radius $r_{\rm
s}=900\,$kpc, ellipticity $e=0.2$ and position angle of 330\,degrees
(measured counterclockwise from north). The cluster centre is marked
by the cross. See the text for details on source positions, redshifts
and sizes. The images have been labelled with upper case letters.  B'
is the predicted counter arc of arc B.}
\label{arcs}
\end{figure*}

In Fig.~\ref{arcs} (left panel) an optical B-R colour image is shown
of the central 1.5$\times$1.5 arcmin of Abell 1835. The image has been
constructed from the 20\,min B-band image shown in Fig.~\ref{B-band}
and a 10\,min R-band image from the same observing run that was also
taken from the CFHT archive. In this greyscale image, blue objects
appear black, and red objects appear white. The contrast has been
stretched that very faint objects can be seen. The two vertical
stripes are due to bleeding columns in the R-band image that
fortunately did not affect the cD galaxy.

In the central part of Fig.~\ref{arcs} (left panel), it can be seen
that only the cD galaxy appears blue.
%The four objects close to the central galaxy
%in Fig.~\ref{B-band} are not blue and have colours consistent with
%being cluster members.
The two arcs to the south east and south west
of the cD galaxy were already pointed out in Fig.~\ref{B-band}. The
larger arc to the south east was discovered by Edge et al.~(in
preparation). The other arc has not been discussed in the literature
so far. No redshift has been determined for either object in the
literature, but their optical appearance and colours indicate that
they are gravitationally lensed images of background galaxies that are
situated at some distance behind the galaxy cluster \citep*[e.g.,
][]{Schneider92}.

In the R-band image (not shown) we find a further pair of very thin
and elongated features parallel to, but 9 arcsec south of the south
east arc (termed C1 and C2 in the following). This pair of arcs is too
faint to be seen in Fig.~\ref{arcs} (left panel). The total length of
the pair is comparable to arc A, but is only barely visible in the
B-band image. If we assume that the very faint B-band brightness of
the pair is due to the redshifting of the Lyman break out of the
B-band filter used (3800-4800\,\AA), we may deduce a redshift of
$z\geq4.0$ for this object pair \citep[e.g., ][]{Madau96}.

The separations of the centres of arcs A and B from the centre of the
cD galaxy are $\theta_{\rm A}\sim 31.1$\,arcsec and $\theta_{\rm
B}\sim 25.0$\,arcsec. In a circularly symmetric gravitational lens,
arcs appear close to the Einstein radius $\theta_{\rm E}$
\begin{equation}
\theta_{\rm E}=\sqrt{\frac{D_{{\rm clus-arc}}}{D_{\rm
clus}\,D_{\rm arc}} \frac{4\,{\rm G}\,M}{c^2}},
\label{einstein}
\end{equation}
where $M$ is the enclosed mass \citep[e.g., ][]{Schneider92}. $D_{\rm
clus}=1.02\,$Gpc, $D_{\rm arc}$ and $D_{\rm clus-arc}$ are the angular
size distances of the cluster, the arc, and between cluster and arc. G
is the gravitational constant and $c$ is the speed of light. For the
case of Abell 1835, \citet{Allen96} calculated the mass in a circular
aperture inside arc A (their Fig.~13) and found a value between
1.4$\times 10^{14}$\,M$_{\odot}$ and 2$\times 10^{14}$\,M$_{\odot}$
for values of the redshift of the arc $0.6<z_{\rm arc}<3.0$. Similar
values follow for arc B.

The NFW mass models in the region of best goodness of fit in
Fig.~\ref{grid} have only little variation of the total mass inside
31.1\,arcsec (153\,kpc): If we use the \citet{Bartelmann96} formulae
for the total mass in a circular NFW profile, the $1\sigma$ formal
confidence interval data ranges from $M_{\rm
X}=1.06\times\,10^{14}\,M_{\odot}$ to $M_{\rm
X}=1.27\times\,10^{14}\,M_{\odot}$ (Fig.~\ref{massprof}). The values
agree with the $1.26^{+0.32}_{-0.17}\times 10^{14}\,M_{\odot}$
determined using {\it ROSAT} data by \citet{Allen98}. The values are
lower, however, than the estimate from the lensing argument assuming
circular symmetry.

From the X-ray image in Fig.~\ref{Image150kpc} we can deduce that in
the inner parts the X-ray isocontours are elliptical with an axis
ratio of $f=0.85$ (Sect.~\ref{axisratio}). This suggests that also the
mass distribution of Abell 1835 is not circularly symmetric. Numerical
simulations by \citet{Bartelmann95} showed that the lensing mass
estimate under the assumption of circular symmetry needs to be
corrected towards lower values. As an average correction factor he
quotes 1.6, but this may vary downwards to unity and upwards to
$\sim$2 from case to case. For the lensing mass estimate the
ellipticity thus could be an important factor. In order to go beyond
the approximation of circular symmetry, we need to use an elliptical
mass model.

The normalized surface mass density $\kappa=\Sigma/\Sigma_{\rm crit}$
of a spherical NFW mass distribution was derived by
\citet{Bartelmann96}. Using the ratio $x=r/r_{\rm s}$ of radius $r$
and scale radius $r_{\rm s}$ this is given by
\begin{equation}
\kappa(x)=2\,\kappa_{\rm s}\,\frac{f(x)}{x^2-1},
\label{kappa_nfw}
\end{equation}
where
\begin{equation}
f(x)=
\left\{
\begin{array}{lc}
1-\frac{2}{\sqrt{x^2-1}}\,{\rm arctan}\,\sqrt{\frac{x-1}{x+1}} &
(x>1)\\
1-\frac{2}{\sqrt{1-x^2}}\,{\rm arctanh}\,\sqrt{\frac{1-x}{1+x}} &
(x<1)\\
0 & (x=1)
\end{array}\right.
\end{equation}
($\kappa\,(1)=\frac{2}{3}\,\kappa_{\rm s}$) and $\kappa_{\rm
s}=\delta_{\rm c}\,\rho_{\rm crit}\,r_{\rm s}\,\Sigma^{-1}_{\rm
crit}$. The critical surface mass density $\Sigma_{\rm crit}$ is given
by
\begin{equation}
\Sigma_{\rm crit}=\frac{c^2}{4\,\pi\,{\rm G}}\,\frac{D_{\rm arc}}{D_{\rm
clus}\,{D_{\rm clus-arc}}},
\end{equation}
with constants as described after eq.~(\ref{einstein}). One can then
generalize eq.~(\ref{kappa_nfw}) for a mass distribution with
elliptical symmetry by making the transformation
\begin{equation}
x\rightarrow x_{\rm
e}=\sqrt{\frac{x_1^2}{(1+\epsilon)^2}+\frac{x_2^2}{(1-\epsilon)^2}}
\end{equation}
\citep*{Schramm90,Schramm94,Kassiola93,Kormann94} where $x_1$ and
$x_2$ are cartesian coordinates and $\epsilon$ is the elliptical
parameter. The elliptical parameter $\epsilon$ is related to the axis
ratio $f$ via $f=\frac{1-\epsilon}{1+\epsilon}$. Using the summation
method developed by \citet{Schramm94} we can then construct
gravitational lens models for the elliptical NFW mass model within the
viable range of parameters from Fig.~\ref{grid}. We use an axis ratio
$f=0.8$ and a position angle of 330 degrees (measured counterclockwise
from north) taken from the optical appearance of the cD galaxy. We do
not use the axis ratio of the X-ray gas because the gas pressure
causes the gas distribution to be more circular than the mass
distribution \cite[e.g., ][]{Buote96}. We fix the centre of the mass
model to the optical centre of the cD galaxy.

%We
%adjust the normalization so that the mass inside a circular aperture
%of 0.153\,Mpc (the radius of arc A) is identical to the mass of the
%corresponding X-ray mass model (Fig.~\ref{massprof}).

Without redshift information about the arc candidates, any
parameterized lens modelling of the system is degenerate with respect
to the mass normalization of the cluster. It is also necessary to make
an assumption about whether to interpret the arcs as single images
which have been elongated by the lens effect, or as fold-arcs
consisting of two mirror images of the same object \citep[e.g.,
][]{Schneider92}. Since we do not detect counter images for arc A, it
is likely that it is a single, elongated image of background
galaxy. We also cannot detect a counter image for the pair C1/C2 in
the CFHT R-frame, which could indicate that they are bright spots in a
common, darker background galaxy that appears elongated by the lens
effect.  On the basis of the archival CFHT data, however, it is
impossible to identify unambiguously which images belong together, or
to rule out the existence of further counter images.

We have first studied the lens effect of an NFW model with
$\sigma=1460\,$km$\,$s$^{-1}$ and $r_{\rm s}=0.9\,$Mpc, which is
situated inside the $1\,\sigma$ confidence region in Fig.~\ref{grid}.
With this mass distribution, arc A can be explained as a single
elongated image of a background galaxy at $z_{\rm A}\sim2.7$. The arc
pair C1/C2 in this model are images of two objects at $z_{\rm
C}\sim5.0$ with a projected physical (unlensed) separation of
$\sim10\,$kpc.  This redshift is consistent with our limit $z_{\rm
C}\geq4.0$ which was obtained from the fact that the pair is not
visible in the CFHT B-frame. An arc at the position of arc B is
predicted for several source redshifts, but the most interesting
interpretation is a background object at $z_{\rm B}\sim2.25$, because
at this redshift the lens model predicts counter images in the form of
a radial arc at the position of the straight blue feature to the north
of the cD (Fig.~\ref{arcs}, left panel; the existence of radial arcs
in NFW profiles was shown by \citealt{Bartelmann96}).

The image configuration predicted by this model is shown in the right
panel of Fig.~\ref{arcs}. Circular Gaussian brightness profiles have
been assumed for all sources with a full width at half maximum of
0.45\,arcsec for source A and 0.25\,arcsec for sources B, C1 and
C2. The contours have been spaced at 10\%, 30\% and 50\% of the
central surface brightness. The assumed source positions on the sky
with respect to the cluster centre in the form $(\Delta x$, $\Delta
y)$ are: source A: ($-5.06$", $-6.20$"), source B: ($0.50$",
$-2.44$"), source C1: ($-7.05$", $-9.03$"), source C2: ($-5.78$",
$-10.22$").

Secondly, since arc A appears roughly symmetric (Fig.~\ref{arcs}, left
panel), we have also investigated the possibility that it is a
fold-arc. This interpretation may be supported by the similarity of
arcs C1 and C2. We find, however, that none of the models inside the
$1\,\sigma$ confidence region in Fig.~\ref{grid} are massive enough to
explain a fold-arc at the radius of arc A (31.1\,arcsec or
153\,kpc). In this case it would be necessary to scale the surface
mass density $\kappa\rightarrow\kappa'=q\times\kappa$ of all models in
the formal $2\,\sigma$ contour by correction factors between $q=1.1$
and $q=1.4$. The redshift of arc A in such models would typically be
$z_{\rm A}\sim2.0$. (The redshift is constrained to be around $z_{\rm
A}\sim2.0$ because it is also necessary to explain arcs B and
C1/C2. Arc B would have a redshift $z_{\rm B}\sim0.9$. C1 and C2 would
be interpreted as images of a source at $z_{\rm C}>4.2$. The radial
feature to the north of the cD could again be interpreted in this
model as a radial arc.)

We have also studied different axis ratios within the context of the
fold-arc model: Lowering the axis ratio reduces the correction factor
$q$ because the ellipticity causes the critical curve to extend
further out. We thus find that elliptical $f\sim0.6$ models within the
$2\sigma$ confidence contour of Fig.~\ref{grid} are able to explain
the CFHT data. The axis ratio should be larger than $f=0.5$, however,
because otherwise it is not possible to have an elongated arc at the
position of arc B (without additional perturbations of the
potential). By using nonsingular isothermal mass profiles
(eq.~(\ref{isoc})) the correction factor $q$ can also be reduced
further.
%We note, however, that all fold-arc models make predictions for
%currently undetected counter images of arcs A, C1 and C2, as well as
%for the radial feature if interpreted as a gravitational arc.

Our analysis clearly favours the first, single-image, interpretation
of arc A, as well as C1/C2, because the fold-arc model predicts
counter-images for arcs A, C1/C2 that are not observed. The available
strong lensing data are explained by a mass model entirely consistent
with the {\it Chandra} temperature profile, indicating that there are
no significant contributions from non-thermal sources of pressure
(e.g., bulk motions, magnetic fields). The fold-arc models require
special tuning of the mass distribution, although it may be possible
to explain the additional mass with systematic uncertainties of our
X-ray mass analysis due to the residual substructure seen in the X-ray
emission, or by a merging subcluster or some other line-of-sight
enhancement present in this region. Nevertheless, the predicted
counter images would have to be found before this option can be
considered further.

We have not attempted any more detailed lens modelling including
sub-halos due to cluster member galaxies because the available data
are of insufficient quality to unambiguously establish the arc
properties and identify image pairs. A more rigorous analysis aiming
to place precise constraints on the redshifts of the arcs and
corresponding arc properties is a subject for future study and will
best be done with the Hubble Space Telescope
\citep{Kneib00,Smith00}. Moreover, spectroscopy will be required in
order to ensure that the arc candidates are indeed galaxies at high
redshift \citep[e.g., ][]{Ebbels98}. A measurement of the redshift of
arc B alone would provide a very strong test of the mass models
discussed here.

\subsection{The X-ray gas mass fraction}

\begin{figure}
\begin{center}
\resizebox{\columnwidth}{!}
{\includegraphics{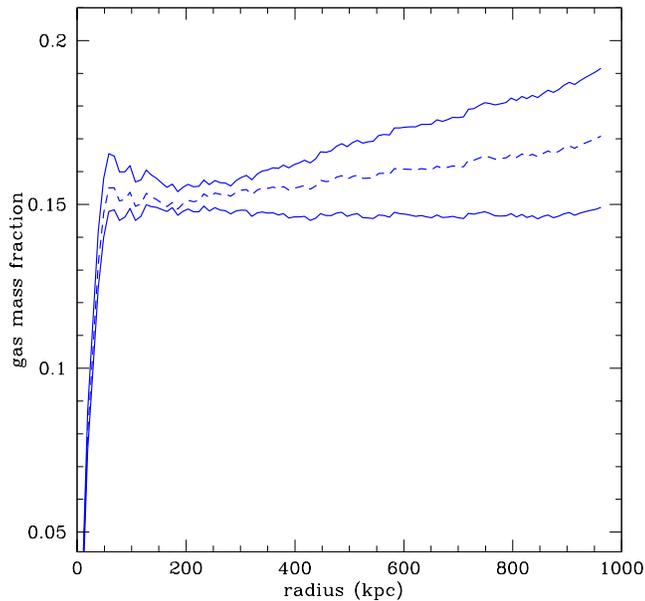}}
\caption{Upper and lower $1\,\sigma$ limits on the gas mass fraction
fraction as a function of radius~$r$. In addition, for reference the
gas mass fraction has been plotted also for the best-fit model $r_{\rm
s}=0.64$\,Mpc, $\sigma=1275\,$km\,$s^{-1}$ (dashed line).}
\label{baryon}
\end{center}
\end{figure}

The X-ray gas-to-total-mass ratio as a function of radius, $f_{\rm
gas}(r)$, determined from the {\it Chandra} data is shown in
Fig.~\ref{baryon}. For this figure, the gas mass fraction was
calculated for all models contained in the $1\,\sigma$ confidence
contour of Fig.~\ref{grid}. The dashed line corresponds to the
best-fit model, the solid lines are the envelopes of all calculated
profiles. The $f_{\rm gas}$ value rises rapidly with increasing radius
within the central $\sim 50$ kpc and then flattens, remaining
approximately constant out to the limits of the data at $r=1.0$ Mpc,
where we measure $f_{\rm gas}=0.171^{+0.020}_{-0.022} h_{\rm
50}^{-1.5}$.

Following the usual arguments, which assume that the properties of
clusters provide a fair sample of those of the Universe as a whole
(e.g., \citealt*{White93,White95,Evrard97,Ettori99}), we may use our
result on the X-ray gas mass fraction in Abell 1835 to estimate the
total matter density in the Universe, $\Omega_{\rm m}$. Assuming that
the luminous baryonic mass in galaxies in Abell 1835 is approximately
one fifth of the X-ray gas mass (e.g., \citealt{White93},
\citealt*{Fukugita98}) and neglecting other possible sources of
baryonic dark matter in the cluster, we obtain $\Omega_{\rm m} =
(\Omega_{\rm b}/1.2 f_{\rm gas})$, where $\Omega_{\rm b}$ is mean
baryon density in the Universe and $h_{50}$ is the Hubble constant in
units of 50\,km\,s$^{-1}$\,Mpc$^{-1}$. For $\Omega_{\rm b}h_{\rm
50}^{2} = 0.0820\pm0.0072$ \citep{Omeara00}, we measure $\Omega_{\rm
m} = 0.40 \pm 0.08 h_{\rm 50}^{-0.5}$.

\section{The properties of the cooling flow}
\label{cflow}

\subsection{Density and Cooling times}
\label{density}

\begin{figure}
\begin{center}
\resizebox{\columnwidth}{!}  {\includegraphics{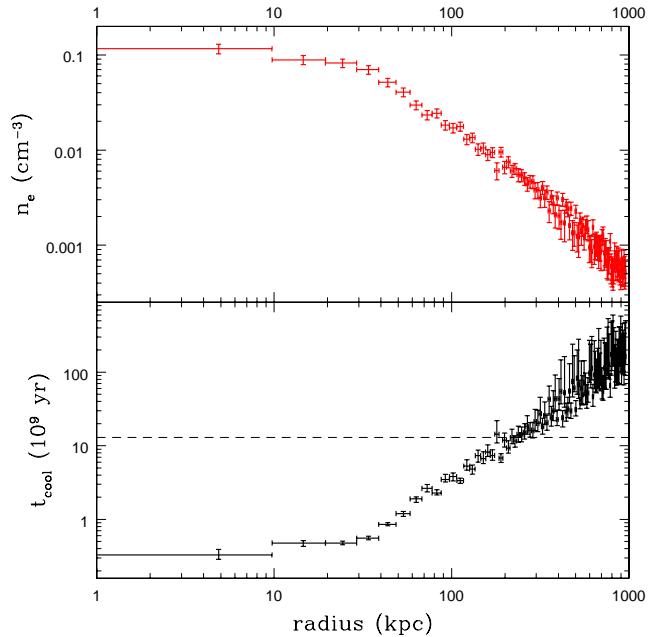}}
\caption{Electron density $n_{\rm e}$ and cooling time $t_{\rm cool}$
as a function of radius $r$. The cooling radius $r_{\rm cool}$ at
which the cooling time profile becomes larger than the age of the
universe $\frac{2}{3}\,$H$_0=13\,$Gyr (dashed line) is $r_{\rm
cool}\sim 230\,$kpc.}
\label{ne_cool}
\end{center}
\end{figure}

Using the density and temperature profiles determined with the
best-fit NFW mass model in Sect.~\ref{massmodel} we can calculate the
cooling time of the intracluster gas (i.e. the time taken for the gas
to cool completely) as a function of radius. The results on the
electron density and cooling time profiles are shown in
Fig.~\ref{ne_cool}. The overall profile of the density data can be
approximated by a $\beta$-model
\begin{equation}
n_{\rm e}(r)\propto \left[1+\left(\frac{r}{r_{\rm c}}\right)^2
\right]^{-\frac{3\,\beta}{2}}
\label{betamodel}
\end{equation}
with a core radius, $r_c = 39\pm3$ kpc, $\beta=0.53\pm0.01$ and a
central density, $n_{\rm e}(0)=10.2\pm0.8 \times10^{-2}$ cm$^{-3}$.

\begin{figure}
\begin{center}
\resizebox{\columnwidth}{!}
{\includegraphics{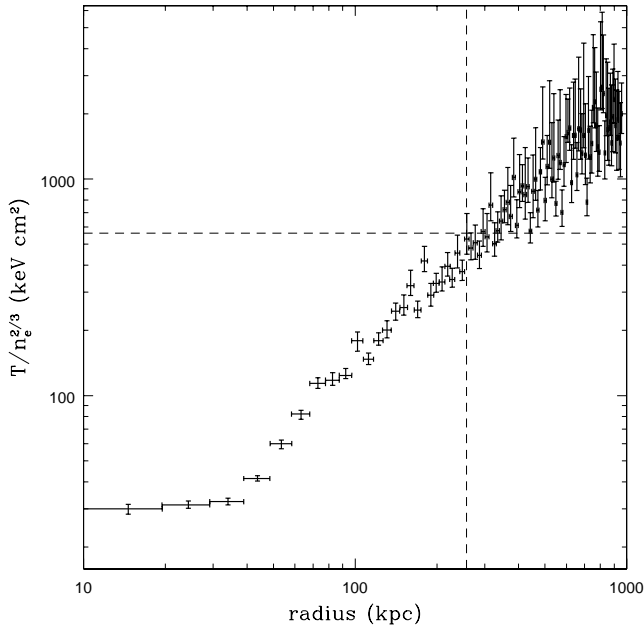}}
\caption{Entropy S as a function of radius. To guide the eye the
entropy value for a temperature of 10\,keV according to the numerical
relation from \protect\citet*{Ponman99} has been plotted with dashed
lines: S=567.0 at 0.1 virial radii, 0.1$\,c\,r_{\rm s}=257$\,kpc
(Sect.~\protect\ref{massmodel}).}
\label{entropy}
\end{center}
\end{figure}

The dashed line in the lower panel of Fig.~\ref{ne_cool} indicates the
age of the universe for our chosen cosmology with
$H_0=50\,$km$\,$s$^{-1}\,$Mpc$^{-1}$. For the assumed Galactic column
density of $2.3\times10^{20}$\,cm$^{-2}$ \citep{Dickey90}, we measure
a central cooling time of $t_{\rm cool} \sim 3{\times}10^8$ yr, and a
cooling radius, at which the cooling time first exceeds the age of the
universe, of $r_{\rm cool} \sim 230$\,kpc. In Fig.~\ref{entropy} we
plot the entropy $S=\frac{T}{n_{\rm e}^{2/3}}$ as a function of
radius, calculated using the temperature and electron density profiles
of the best-fit model.

The fact that cooling times in the centres of cooling flow clusters
are comparable to or even significantly less than the age of the
universe is well known \citep{Lea73,Silk76,Cowie77,Fabian77} and was
an important motivation for the development of cooling flow models. In
the next sections we explore the properties of such a cooling flow
model for Abell 1835.

\subsection{Standard cooling flow modelling and image analysis}
\label{coolingflow}

In this section we report on the results from fitting cooling flow
models to the spectra extracted from full circles centred on the
geometrical cluster centre in the 0.5-7.0\,keV band. In the standard
cooling flow picture this yields the integrated mass deposition rate
of the matter that cools out of the flow within a given radius.

We model the cooling flow using the RJCOOL emission model constructed
according to the prescription in \citet{Johnstone92} and include
intrinsic absorption that may be acting on the cooling flow in the
core of the cluster with a ZPHABS absorbing model
\citep{Balucinska-Church92} at the redshift of the cluster. We use the
MEKAL plasma emission model \citep{Kaastra93,Liedahl95} to model the
gravitational work done on the gas as well as emission of the
intracluster gas outside the cooling flow:
\begin{eqnarray}
{\rm MODEL}_2 & = & {\rm PHABS}_{\rm \,gal}\,\times \left[{\rm
MEKAL}\,(T_{\rm high};Z;K) \frac{}{} \right.\nonumber \\
& & \left. +\,{\rm ZPHABS}\, ( \Delta N_{\rm H} )\,\times\, {\rm
RJCOOL}\, (\dot{M}) \right].
\label{rjcool}
\end{eqnarray}
The free parameters are given in brackets. The overall absorption in
this model is fixed at the Galactic value \citep{Dickey90}. The
intrinsic absorption $\Delta N_{\rm H}$ acts only on the cooling flow
emission.  The normalization of the cooling flow model, $\dot{M}$,
measures the rate of gas cooling out of the flow. The other parameters
of the cooling flow model, namely the metallicity $Z$ and upper
temperature $T_{\rm high}$, are fixed to the values of the MEKAL
component. This model is identical to model C in \citet{Allen00}. It
is a multi-phase model in the sense that at each radius all
temperatures from the upper temperature down to the lowest temperature
are present.

\begin{figure}
\begin{center}
\rotatebox{270}{
\resizebox{!}{\columnwidth}{\includegraphics{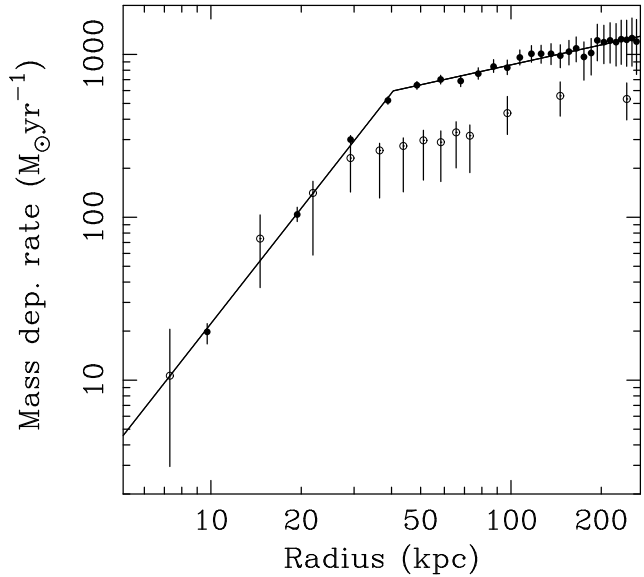}}
}
\caption{Integrated mass deposition rate $\dot{\rm M}$ as a function
of radius $r$ determined spectrally (open circles) and from the image
analysis (filled circles). The solid line is a broken power-law fit to
the image analysis profile with a break at $r_{\rm break}=42$\,kpc and
exponents of 2.4 before and 0.4 after the break.}
\label{mdot}
\end{center}
\end{figure}

In Table~\ref{rjcool_tab_110} the results of the model fits are given
for outer radii from 5\,arcsec (25\,kpc) to 50\,arcsec (250\,kpc). In
Fig.~\ref{mdot} we have plotted with open circles the mass deposition
results with an even finer spacing of the extraction circles. All of
the fits are statistically acceptable, with reduced $\chi^{2}$ values
around or slightly above unity. There is no significant improvement in
$\chi^2$ over the simpler isothermal
models~eq.~(\ref{isothermal}). The {\it Chandra} data for Abell 1835
cannot discriminate between single phase and multi-phase models.

Also shown in Fig.~\ref{mdot} (solid circles) is the mass deposition
rate that may be deduced from the X-ray luminosity distribution
\citep[e.g., ][]{White97} using our best-fit mass model from
Sect.~\ref{massmodel}. It can be seen that the results from the
spectral modelling and from the image analysis agree rather well out
to 30 kpc, where the spectral mass deposition profile has a break at a
mass deposition rate of 231$^{+79}_{-53}\,M_{\odot}\,$yr$^{-1}$. The
profile from the image analysis then itself has a break around
40\,kpc. The solid line corresponds to a broken power-law fit to the
mass deposition profile from the image analysis. Both profiles rise
with a similar slope after their respective breaks, but disagree with
respect to the amplitude. The spectral mass deposition profile rises
up to a value of $\sim\,500\,M_{\odot}\,$yr$^{-1}$ within 250\,kpc,
whereas the profile from the image analysis rises to values of
1250\,$M_{\odot}\,$yr$^{-1}$, which is consistent with the value from
{\it ROSAT} \citep{Allen00}.

\begin{table*}
\caption{Cooling flow model fits (eq.~(\ref{rjcool})) to spectra from
circular regions with radii $r$. We list the best-fit values for the
model parameters mass deposition rate $\dot{M}$, upper temperature
$T_{\rm high}$, metallicity $Z$ and intrinsic equivalent hydrogen
column density $\Delta\,N_{\rm H}$. We also list the $\chi^2$ value,
the number of degrees of freedom (DOF) and the reduced $\bar{\chi}^2$
value (see Table~\ref{mekal_tab_110}).}
\label{rjcool_tab_110}
\begin{center}
\begin{minipage}{140mm}
\begin{center}
\begin{tabular}{@{}rrrrrlll@{}}
$r$ (kpc) & $\dot{M}$ ($M_{\odot}\,$yr$^{-1}$) & kT$_{\rm high}$
(keV) & $Z$ (solar) & $\Delta N_{\rm H}$ (10$^{22}$\,cm$^{-2}$) & $\chi^2$ &
DOF & $\bar{\chi}^2$\\
\hline
25 & 162.4$^{+70.0}_{-57.9}$ & 4.8$^{+0.9}_{-0.5}$ & 0.23$^{+0.08}_{-0.07}$ & 0.20$^{+0.10}_{-0.05}$ & 144.8 & 121 & 1.20\\
50 & 284.6$^{+68.5}_{-93.2}$ & 4.9$^{+0.2}_{-0.3}$ & 0.27$^{+0.05}_{-0.04}$ & 0.31$^{+0.13}_{-0.06}$ & 240.3 & 187 & 1.29\\
100 & 383.7$^{+102.2}_{-99.7}$ & 6.3$^{+0.4}_{-0.3}$ & 0.25$^{+0.04}_{-0.04}$ & 0.26$^{+0.09}_{-0.06}$ & 291.4 & 255 & 1.14\\
150 & 489.9$^{+106.6}_{-122.1}$ & 7.1$^{+0.4}_{-0.4}$ & 0.26$^{+0.04}_{-0.03}$ & 0.20$^{+0.06}_{-0.03}$ & 333.7 & 285 & 1.17\\
250 & 468.7$^{+119.5}_{-119.9}$ & 7.5$^{+0.4}_{-0.3}$ & 0.28$^{+0.03}_{-0.04}$ & 0.23$^{+0.07}_{-0.05}$ & 389.4 & 323 & 1.21\\
%375 & 594.2$^{+130.8}_{-115.4}$ & 8.0$^{+0.4}_{-0.3}$ & 0.28$^{+0.03}_{-0.03}$ & 0.23$^{+0.05}_{-0.04}$ & 381.2 & 343 & 1.11\\
%500 & 663.6$^{+131.6}_{-133.5}$ & 8.6$^{+0.4}_{-0.4}$ & 0.27$^{+0.03}_{-0.03}$ & 0.19$^{+0.05}_{-0.03}$ & 367.0 & 355 & 1.03\\
%750 & 693.8$^{+141.2}_{-151.4}$ & 9.2$^{+0.6}_{-0.5}$ & 0.28$^{+0.03}_{-0.04}$ & 0.17$^{+0.03}_{-0.03}$ & 401.4 & 377 & 1.06
\end{tabular}
\end{center}
\end{minipage}
\end{center}
\end{table*}

The intrinsic absorption $\Delta N_{\rm H}$ tabulated in
Table~\ref{rjcool_tab_110} acting on the cooling flow emission rises
from a mean equivalent hydrogen column density of $\Delta\,N_{\rm
H}\sim\,0.2\,\times\,10^{22}\,$cm$^{-2}$ at a radius of 50\,arcsec
(250\,kpc) by a factor of $\sim\,1.5$ when we look deeper into the
cluster core. The statistical error bars on the intrinsic absorption
are large, however. The temperature of the MEKAL component (which is
linked to the upper temperature of the cooling gas) rises from about
5\,keV in the centre to 9\,keV within a circular aperture of 750\,kpc
radius.

\subsection{Interpretation}

\subsubsection{A young central cooling flow?}

Standard comoving multiphase cooling flow models \citep[e.g.,
][]{Nulsen86} typically predict the profile of the mass deposition
rate as a function of radius to be a power law. A break in the profile
of the mass deposition rate, as seen in Fig.~\ref{mdot}, may suggest
that the flow mechanism is changing around the break radius.
\citet{Allen01b} interpret the presence of such breaks as an indicator
of the age of the cooling flow, which they suggest is likely to be
closely related to the time since the central regions of the cluster
settled down following the last significant disruption by merger
events or AGN activity. Given the cooling time at a radius of 30\,kpc
(Fig.~\ref{ne_cool}), this would then limit the age of the cooling flow
in Abell 1835 to be $\lesssim 6{\times}10^8$\,years.

The natural question which must be asked is whether such a young age
for the cooling flow in Abell 1835 is plausible. Given the clear
substructure observed in the core of the cluster
(Sect.~\ref{imaging}), some disturbance in the recent past seems
likely. As discussed in Sect.~\ref{imaging}, the small emission front
seen in Fig.~\ref{Image150kpc} to the south-east of the cluster core
may have been caused by a cool, infalling gas cloud~\citep[e.g., see
the work of~][]{Markevitch00}. A 31.3 mJy FIRST radio source
\citep{Becker00} is located in the cD galaxy. Although this source is
relatively quiescent at present, it may have been much more powerful
in the recent past and injected energy and momentum into the
intracluster medium (possible problems with radio source heating are
discussed by \citealt{David01} and \citealt{Fabian01}). One should
also recall that within the context of structure formation models, a
cluster as hot and massive as Abell 1835, observed at a redshift $z
\sim 0.25$, is highly unlikely to have formed more than a few Gyr
before.

If the cluster remains undisturbed, the results on the cooling times
and the mass deposition rate in Figs.~\ref{ne_cool} and~\ref{mdot}
suggest that the core of the cluster will develop into a very massive
cooling flow within a few Gyr. It is also possible that the cluster
contained a large cooling flow before its most recent
disturbance. Note that similar studies of the cooling flows in the
Abell 2390 and 3C295 clusters indicate older ages of 2-3 and 1-2 Gyr,
respectively.

We note that the fact that the break appears at a slightly smaller
radius in the spectral analysis than the image analysis could simply
be a consequence of the fact that although the gas starts to cool
appreciably around 40\,kpc, it has only had sufficient time to cool
completely within $r \sim 30$\,kpc. Note also that given the complex
central morphology in Fig.~\ref{Image150kpc}, simple,
spherically-symmetric cooling-flow models are unlikely to be
applicable in detail.
% and the cooling may be concentrated within the
%coolest, brightest blobs.

As discussed by \citet{Allen01a}, in cases where the temperature of
the cluster gas continues to rise beyond the outer edge of the cooling
flow, spectral-measurements of the mass deposition rate covering
regions larger than the cooling flow and using models similar to
eq.~(\ref{rjcool}), are likely to overestimate the true mass
deposition rate. This is because the presence of relatively cool,
ambient spectral components will tend to be modelled as part of a
cooling flow. This situation is likely to have affected previous ASCA
studies of the integrated cluster spectrum for Abell 1835
\citep{Allen96,Allen00}. We also note that the application of such
models to large spatial regions may lead to the detection of
low-temperature `cut-offs' in the cooling flow spectra, which would
simply be due to the fact that complete cooling occurs only within the
innermost cooling flow proper.

\subsubsection{XMM observations of Abell 1835}
\label{xmm}

In a recent paper \citet{Peterson01} present new results on grating
spectra of Abell 1835 taken with the Reflection Grating Spectrometer
(RGS) aboard the {\it XMM-Newton} X-ray observatory. They find no
evidence for line emission from gas below 2.7\,keV and, neglecting
intrinsic absorption, place a 90 per cent confidence upper limit on
the mass deposition rate from the cooling flow of $200\,\Msunpyr$ for
$H_0=70\kmpspMpc$, $\Omega_{\rm m}=0.3$ and $\Lambda=0.7$. In fact,
RGS has not found convincing evidence for a cooling flow in any galaxy
cluster yet. Due to the inverse square dependence of the mass
deposition rate on the luminosity distance, the RGS limit translates
into $315\,\Msunpyr$ for the $H_0=50\kmpspMpc$ Einstein-de Sitter
cosmology used here.

Overall, the RGS result for Abell 1835 implies a range of temperatures
from 9 to 3~keV within the inner 0.5 arcmin radius (150\,kpc), with
the emission measures of that gas roughly matching a cooling
flow. This is in agreement with our findings which show a radial
gradient in the ambient gas temperature from ${\sim}8$\,keV at
150\,kpc to ${\sim}4\,$keV in the cluster core. In addition, we
find evidence for complete cooling within the inner 30\,kpc.

The RGS limit on the mass deposition is not easily reconciled with the
large (${\dot M} \gtsim 1000\, $\Msunpyr) mass deposition rates
inferred from previous studies of the integrated cluster spectrum
observed with ASCA and ROSAT. Within the context of the young cooling
flow model discussed above, however, the RGS results can be more
easily accommodated. If one also accounts for the effects of intrinsic
absorption, as required by the {\it Chandra} data, the apparent RGS
limit of $315\,M_{\odot}\,$yr$^{-1}$ can be increased and is
consistent with the {\it Chandra} result of
231$^{+79}_{-53}\,M_{\odot}\,$yr$^{-1}$.

We note that the revised mass deposition rate of $\sim 200-300$
\Msunpyr within $r \sim 30$ kpc in Abell 1835 is comparable to the
optical star formation rate (normalized to the inferred number of A
stars within the same region; \citealt{Allen95,Crawford99}). Thus, it
appears that a significant fraction of the material being deposited by
the cooling flow within the central 30 kpc of Abell 1835 forms stars
with a relatively normal IMF. \citet{Edge01} also reports the
detection of a substantial molecular gas mass ($\sim 2 \times
10^{11}$\Msun) within the central regions of the cluster.

\subsubsection{The origin of the temperature profile}

The temperature profile in a cluster core is a result of the
thermodynamic history of the intracluster gas. {\it Chandra} enables
us for the first time to compare in detail the observed temperature
profiles in clusters with those predicted by numerical studies of
galaxy cluster formation
\citep*[e.g.,][]{Eke98,Frenk99,Lewis00,Pearce00}. In the studies by
\citet{Eke98} and \citet{Frenk99}, the temperature profiles of massive
galaxy clusters (determined with smooth particle hydrodynamics,
without including the effects of cooling), are often observed to drop
slightly towards to core ($r \lesssim 100$ kpc), then remain roughly
constant on a plateau to $r \sim 1$ Mpc, before dropping again at
larger radii.
%The central temperature drop is not 
%well resolved in these since the simulations are typically valid only
%above scales of
%$\sim 30$\,kpc. 

\citet{Pearce00} present temperature profiles for a sample of 20
simulated clusters, both with and without the effects of radiative
cooling. For their largest cluster, the results obtained without
including the effects of cooling are similar to those of \citet{Eke98}
and \citet{Frenk99}.  However, when cooling is included, these authors
find a ``precipitous decline in temperature'' in their largest cluster
which they deem evidence for a cooling flow. The temperature profile
for this cluster appears similar to that observed for Abell 1835
(Fig.~\ref{kT}). The work of Pearce et al. therefore suggests that the
sharp temperature drops in the cores of cooling flow clusters
\citep[see also][]{Allen01a,David01} could be due to the effects of
radiative cooling over the lifetime of the cluster. Since in other
simulations with cooling \citep[e.g., ][]{Lewis00} the temperature
profile seems to keep rising with decreasing radius we note, however,
that the effect of cooling in numerical simulations on the cluster
temperature profiles does not seem to be entirely clear yet.

\subsubsection{A massive cooling flow?}

In this section we address whether an old ($\sim$5\,Gyr), very
massive ($\dot{M}\sim 1000\,M_{\odot}$\,yr$^{-1}$) cooling flow,
encompassing the entire central 150~kpc radius, can exist in Abell
1835. In this case, as mentioned in Sect.~\ref{xmm}, our problem is to
reconcile the large mass flow rate with the relatively small amount of
X-ray line emission observed with RGS.
\citet{Peterson01} and \citet{Fabian01} discuss a number of processes
that may be relevant, two of which are investigated here.

The first model invokes an inhomogeneous distribution of metals that
can be entered into XSPEC using the following prescription:
\begin{eqnarray}
\label{andy_metal}
\lefteqn{{\rm MODEL}_3 = {\rm PHABS}_{\rm gal}\times}\\
&& \left[{\rm MEKAL} \left(T_{\rm high}; Z; K\right)+{\rm
ZPHABS}\left({\Delta}N_{\rm H}\right)\times\right. \nonumber\\
&&\left.\left({\rm RJCOOL}_1 \left(\dot{M};Z=0\right) + {\rm RJCOOL}_2
\left(0.1\,\dot{M};10{\times}Z\right)\right)\right] \nonumber
\end{eqnarray}
where the free parameters (given in brackets) are the same as in
eq.~(\ref{rjcool}), but with the mass deposition rate distributed
between two cooling flow components of which one has zero metallicity
and the other all the heavy elements. This model describes a situation
where most of the heavy elements are confined to small regions, so
that the bulk of the cooling flow is extremely metal poor. The ratio
of the mass of the metal poor and metal rich regions was fixed as
10:1. We have fitted this model to the {\it Chandra} spectrum
extracted from a circle with a radius of 30\,arcsec (150\,kpc). The
best-fit parameters are given in Table~\ref{alter}. The total mass
deposition rate is 1070\,$M_{\odot}$\,yr$^{-1}$. Although the mass
deposition rate is large, however, the emission lines due to cool gas
are suppressed.  The Fe\,XVII line at 15.01{\AA}, for example, is
suppressed by about a factor 4. However, the Fe XXII and Fe XXIII
complex around 12.2\,{\AA}, which is strongly constrained by the RGS
data, is only suppressed by a factor 2 by this model. This appears to
be inconsistent with the RGS result.
\begin{table*}
\caption{Best-fit parameters for possible massive cooling flow
models. The upper limit on the metallicity in the first model was
fixed to the software limit $Z=5$ in the high metallicity component.}
\label{alter}
\begin{tabular}{llllllccc}
model &T$_{\rm high}$ (keV) & T$_{\rm low}$ (keV) &$Z$ (solar) &
$\Delta$N$_{\rm H}\,(10^{22}\,{\rm cm}^{-2})$ &
$\dot{M}\,(M_{\odot}\,{\rm yr}^{-1})$ & $\chi^2$ & DOF &
$\bar{\chi}^2$\\
\hline
inhomogeneous metallicities & 11.0$^{+2.3}_{-2.1}$ & - &
$0.50_{-0.08}$ &
0.08$^{+0.02}_{-0.01}$ & 975.0$^{+170.0}_{-148.0}$ & 322.3 & 285 &
1.13\\ % (10:1)
absorbed cool gas & 9.9$^{+7.1}_{-2.0}$ & 2.6$^{+0.7}_{-0.4}$ &
0.29$^{+0.04}_{-0.04}$ & 0.51$^{+0.15}_{-0.13}$ &
935.0$^{+251.0}_{-187.0}$ & 331.7 & 284 & 1.16
\end{tabular}
\end{table*}

The second model incorporates absorption that only acts on the gas
below a certain temperature threshold and can be entered into XSPEC
using the following prescription:
\begin{eqnarray}
\lefteqn{{\rm MODEL}_4 = {\rm PHABS}_{\rm gal}\times}\\
&& \left[{\rm MEKAL} \left(T_{\rm high}; Z; K\right)+{\rm RJCOOL}_1
\left(\dot{M};T_{\rm low}{\rightarrow}T_{\,\rm high}\right) +
\nonumber\right.\\
&&\left.{\rm ZPHABS}\left({\Delta}N_{\rm H}\right)\times\left({\rm
RJCOOL}_2 \left(\dot{M};0\,{\rm keV}{\rightarrow}T_{\,\rm low}\right)
\right) \right]. \nonumber
\end{eqnarray}
Again, the model is very similar to the one presented in
eq.~(\ref{rjcool}). The best-fit values for the spectrum extracted
from a circle with a radius of 30\,arcsec (150\,kpc) are also given in
Table~\ref{alter}. The emission measure of gas below 2.7\,keV is
suppressed in the same way as the previous model, but this model also
fails to adequately suppress the Fe XXII and Fe XXIII emission around
12.2\,{\AA}.
%It is remarkable that the
%temperature below which the gas is absorbed comes out at similar to
%the cut-off value found by \citet{Peterson01}. This was not a
%constraint of the model.

Although both models have significantly reduced emission from emission
lines by cool ($\lesssim$\,2.7 keV) gas, their emission measure from
gas at intermediate temperatures does not seem to be consistent with
the RGS results.

\section{Conclusions}
\label{conclusions}

We have presented the analysis of 30\,ksec of {\it Chandra} ACIS-S3
observations of the galaxy cluster Abell 1835, of which 20\,ksec were
usable for spectroscopic analysis. Overall, the X-ray image shows a
relaxed looking galaxy cluster. The X-ray isocontours in the centre
are elliptical with an axis ratio of 0.85. In the inner 30\,kpc
radius, however, we detect clear substructure in the image.

Using isothermal fits to spectra from annular regions around the
cluster centre we have measured the temperature, metallicity and
photoelectric absorption of the cluster gas as a function of distance
from the cluster centre. The temperature analysis reveals a steep drop
in the temperature from 12\,keV in the outer regions to only 4\,keV in
the cluster core. We also find evidence for a rise in the
photoelectric absorption towards the cluster core, but no indication
of a change in the metallicity with radius.

The {\it Chandra} data provide tight constraints on the gravitational
potential of the cluster which can be parameterized by an NFW mass
model. The best-fit NFW model has a scale radius $r_{\rm
s}=0.64^{+0.21}_{-0.18}$\,Mpc, an effective velocity dispersion
$\sigma=1275\pm160\,$km\,s$^{-1}$, a concentration parameter
$c=4.0^{+0.54}_{-0.64}$ and a total mass
$M=9{\times}10^{14}\,M_{\odot}$ within 1\,Mpc. We measure the X-ray
gas mass fraction of Abell 1835 as a function of radius and use this
to deduce a cosmic matter density $\Omega_{\rm
m}=0.40\pm0.09\,h_{50}^{-0.5}$.

The projected mass within a radius of $\sim$150\,kpc implied by the
presence of gravitationally lensed arcs in the cluster is in good
agreement with the mass models preferred by the {\it Chandra} data.
This rules out any significant contributions from non-thermal sources
of X-ray gas pressure (e.g., bulk motions, magnetic fields).

The detection of the Sunyaev-Zeldovich (SZ) effect \citep{Sunyaev72}
in Abell 1835 was recently reported by \citet{Mauskopf00}. The
temperature (Fig.~\ref{kt_deprojected}) and electron density
(Fig.~\ref{ne_cool}, upper panel) profiles determined in this paper
can be used to predict the cosmology-dependent SZ effect in the
cluster. This calculation and the cosmological implications of the
SZ-detection by \citet{Mauskopf00} are described elsewhere (Schmidt \&
Allen 2001, in preparation).

%The results on the
%X-ray gas density and temperature profiles are consistent with the
%comptonization parameter determined from observations of the
%Sunyaev-Zeldovich effect by \citet{Mauskopf00}.

The {\it Chandra} data imply a radiative cooling time for the gas in
the centre of Abell 1835 of about $3{\times}10^{8}$\,yr. By fitting
cooling flow models to spectra from circular regions around the
cluster centre, we measure a mass deposition profile that can be
represented by a broken power law with a break at a radius of 30\,kpc,
within which the integrated mass deposition rate is
$231^{+79}_{-53}\,M_{\odot}$\,yr$^{-1}$. We observe a similar
behaviour in the mass deposition profile that can be derived from an
image deprojection analysis, although the break in this case appears
at a slightly larger radius $r\sim$40\,kpc. The spectral and the
imaging mass deposition profiles agree within $r\sim$30\,kpc. If we
associate the cooling time of the X-ray gas at the break radius with
the age of the cooling flow \citep{Allen01b}, we obtain an age of
$\sim 6{\times}10^{8}$\,yr.

Recent results from the {\it XMM-Newton} X-ray observatory
\citep{Peterson01} limit the mass deposition rate in Abell 1835 to
315\,$M_{\odot}\,$yr$^{-1}$, neglecting intrinsic absorption. This
limit appears to be consistent with our result of having a young
cooling flow with a mass deposition rate
$\dot{M}=231^{+79}_{-53}\,M_{\odot}\,$yr$^{-1}$ in the centre of Abell
1835, particularly if one also includes the effects of intrinsic
absorption as required by the {\it Chandra} data. The young cooling
flow scenario is supported by the detection of substructure in the
core, which shows that this region has not completely relaxed. Strong
signs of star formation have been detected in the inner 30\,kpc with
optical spectroscopy \citep{Allen95} that could be fuelled by the
inflowing gas.

\section*{Acknowledgments}

We thank the {\it Chandra} team for the X-ray data. We thank Roderick
Johnstone for providing his script to average response matrices for
extended sources. Stefano Ettori is thanked for software to generate
the exposure map and to extract the surface brightness profile. We
also thank the anonymous referee for a careful reading of the
manuscript. This research has made use of data obtained from the
Canadian France Hawaii Telescope, which is operated by the National
Research Council of Canada, the Centre National de la Recherche
Scientifique of France and the University of Hawaii. We also thank the
observers that proposed and obtained the CFHT data used in this study,
identified as Soucail and Picat in the image headers.  SWA and ACF
acknowledge support by the Royal Society.

%\appendix

%\section[]{}

%
%\newpage
%
%\begin{figure}
%\caption{Figure in Appendix}
%\label{appenfig}
%\end{figure}

\bsp

\label{lastpage}

\end{document}